\documentclass[times,twocolumn,final]{elsarticle}
\usepackage{amsmath,amsfonts,bm}

\def\eqref#1{equation~\ref{#1}}
\def\1{\bm{1}}

\DeclareMathAlphabet{\mathsfit}{\encodingdefault}{\sfdefault}{m}{sl}
\SetMathAlphabet{\mathsfit}{bold}{\encodingdefault}{\sfdefault}{bx}{n}

\usepackage{amssymb}
\usepackage{multirow,setspace,verbatim,amsfonts,amsmath,amsbsy,epsfig,url}

\usepackage{gensymb}
\usepackage{booktabs}

\usepackage{medima}
\usepackage{framed}

\usepackage{amsmath,amsfonts}
\usepackage{latexsym}

\usepackage{hyperref}
\usepackage{url}

\usepackage{array,multirow}
\newcolumntype{P}[1]{>{\centering\arraybackslash}p{#1}}
\newcolumntype{M}[1]{>{\centering\arraybackslash}m{#1}}
\usepackage{mathrsfs} 
\usepackage{algorithm}
\usepackage{algpseudocode}
\usepackage{booktabs}
\usepackage{wrapfig}
\usepackage{graphicx}
\usepackage{caption}
 \usepackage{amssymb}
 
\usepackage{xcolor}

\definecolor{newcolor}{rgb}{.8,.349,.1}
\newcommand{\m}{{\boldsymbol m}}
\newcommand{\x}{{\boldsymbol x}}

\journal{Medical Image Analysis}

\begin{document}

\verso{B. Kim and Y. Oh \textit{et~al.}}

\begin{frontmatter}

\title{C-DARL: Contrastive diffusion adversarial representation learning for label-free blood vessel segmentation}
% \tnotetext[]{This paper extends the work \citep{kim2022diffusion} presented at the Eleventh International Conference on Learning Representations (ICLR) 2023.}
\author[1]{Boah \snm{Kim}\corref{coa}}
\cortext[coa]{Co-first authors.}
\author[2]{Yujin \snm{Oh}\corref{coa}}
\author[1]{Bradford J. \snm{Wood}}
\author[1]{Ronald M. \snm{Summers}\corref{cor1}}
\cortext[cor1]{Co-corresponding authors.}
\ead{rms@nih.gov}
\author[2]{Jong Chul \snm{Ye}\corref{cor1}}
\ead{jong.ye@kaist.ac.kr}
  
\address[1]{Radiology and Imaging Sciences, National Institutes of Health Clinical Center, Bethesda, Maryland, USA}
\address[2]{Kim Jaechul Graduate School of AI, Korea Advanced Institute of Science \& Technology (KAIST),
Daejeon, Republic of Korea}

% \received{1 May 2013}
% \finalform{10 May 2013}
% \accepted{13 May 2013}
% \availableonline{15 May 2013}
% \communicated{S. Sarkar}

\begin{abstract}
%%% The abstract should be no longer than 200 words.
Blood vessel segmentation in medical imaging is one of the essential steps for vascular disease diagnosis and interventional planning in a broad spectrum of clinical scenarios in image-based medicine and interventional medicine. Unfortunately, manual annotation of the vessel masks is challenging and resource-intensive due to subtle branches and complex structures. To overcome this issue, this paper presents a self-supervised vessel segmentation method, dubbed the contrastive diffusion adversarial representation learning (C-DARL) model. Our model is composed of a diffusion module and a generation module that learns the distribution of multi-domain blood vessel data by generating synthetic vessel images from diffusion latent. Moreover, we employ contrastive learning through a mask-based contrastive loss so that the model can learn more realistic vessel representations. To validate the efficacy, C-DARL is trained using various vessel datasets, including coronary angiograms, abdominal digital subtraction angiograms, and retinal imaging. Experimental results confirm that our model achieves performance improvement over baseline methods with noise robustness, suggesting the effectiveness of C-DARL for vessel segmentation.
\footnote[1]{\fntext[1]{This paper extends the work \citep{kim2022diffusion} presented at the Eleventh International Conference on Learning Representations (ICLR) 2023.}}
%%%%
\end{abstract}


\end{frontmatter}

% \linenumbers

\section{Introduction} 
\begin{figure}[!t]
\centering
\includegraphics[width=\linewidth]{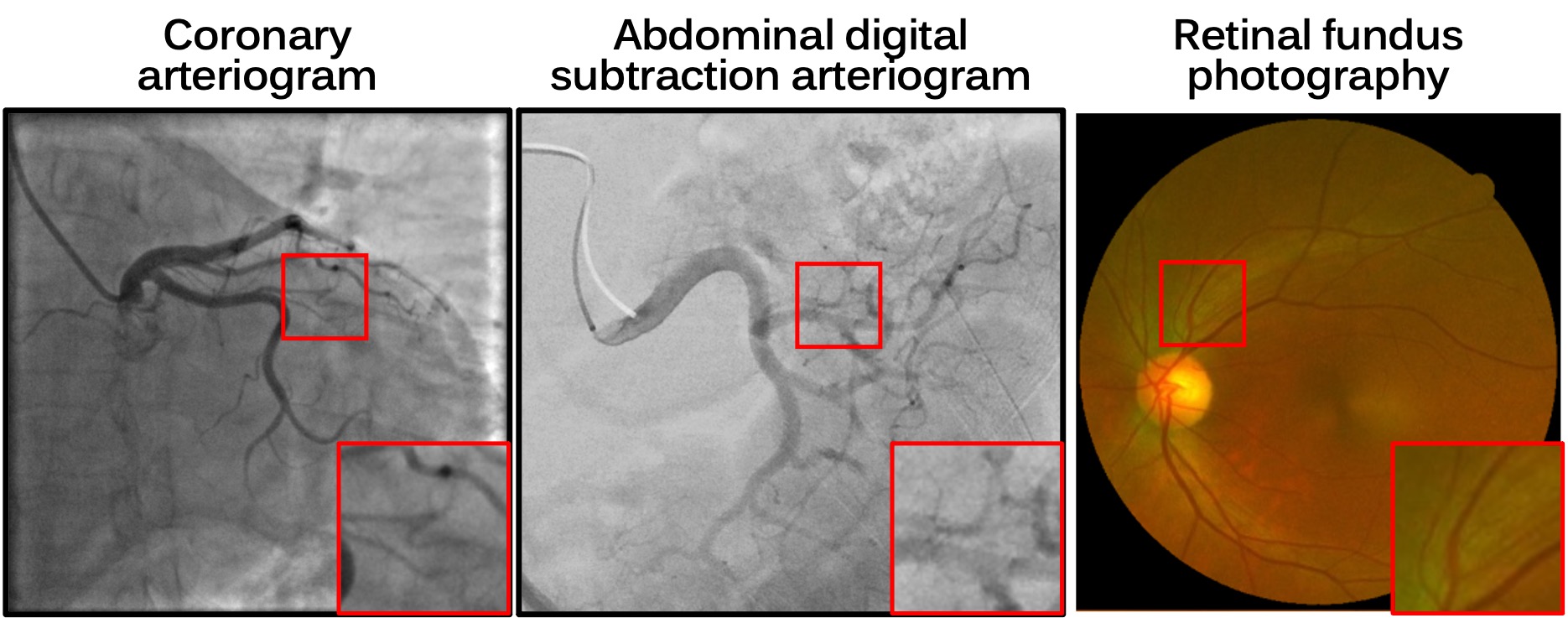}
\caption{Examples of various blood vessel image domains. Red boxes show magnified vessel structures.}
\label{fig:intro}
\end{figure}

Angiography is an invasive or non-invasive exam to visualize blood vessels towards diagnosis or treatment of a wide variety of diseases that impact vascular structures, or where vascular maps provide the roadmap to delivery of therapeutics.
% to diagnose and treat vascular diseases. 
For example, to plan therapy and accurately deliver drugs and devices in minimally invasive image-guided therapies, identification, characterization, and quantification of the blood vessels and their branches is a foundational element towards blood flow, endothelial pathology, landmarks, reference points and roadmaps towards tumors or target anatomy \citep{dehkordi2011review}.
% important since the vascular structures provide blood information and landmarks to the lesion \citep{dehkordi2011review}. 
As manual annotation of vessel masks is time-consuming due to tiny and low-contrast vessel branches (see Figure~\ref{fig:intro}), automatic vessel segmentation methods have been extensively studied to enhance efficiencies and to facilitate large data for training \citep{delibasis2010automatic, jiang2019automatic, wu2019u}.

%%%%%% Vessel segmentation methods
Classical rule-based vessel segmentation methods utilize various features of vessel images such as geometric models, ordered region growing, and vessel intensity distributions \citep{lesage2009review, taghizadeh2014local, zhao2019segmentation}. However, these approaches require complicated preprocessing and manual refinement, posing resource barriers and challenges to % in 
practical clinical deployment. On the other hand, recent learning-based techniques \citep{nasr2016vessel, fan2018multichannel, wu2019u}, which segment blood vessels through neural networks, can generate outputs in real-time, but they require the supervision of large amounts of annotated data.

%%%%%% Motivation of our method
Recently, self-supervised learning methods, which do not require ground-truth vessel masks when training networks, have been extensively studied. For example, \cite{ma2021self} presented an adversarial vessel segmentation method using fractal masks. \cite{kim2022diffusion} proposes a diffusion adversarial representation learning model (DARL) that combines the diffusion model and the adversarial model. Specifically, the DARL model learns the distribution of background images using the diffusion denoising probabilistic model (DDPM) \citep{ho2020denoising} so that the vessel structures can be easily discerned in the latent. Accordingly, a subsequent generation module can extract foreground vessel regions. Although this method provides a high-quality vessel segmentation map through single-step inference {while alleviating the network training complexity compared to \cite{ma2021self}}, one of the main limitations of DARL is that it uses both the pre-contrast background and angiography images for network training, which may limit its use in various clinical applications that do not normally provide similar background images (eg., retinal fundus images). Future training with digital subtraction and raw images might reduce this training variability.

%%%%%% Brief description of our method
This paper is to extend the concept of DARL and present a label-free vessel segmentation method that can utilize a variety of blood vessel images in training the model, enabling its generalizability as a vessel segmentator in a variety of clinical applications. Specifically, we design a model that basically follows the overall structure of DARL consisting of the diffusion and generation modules, with customizations. One of the key improvements over DARL comes from our observation that we can still obtain a semantically meaningful segmentation map by omitting the background image input path. 

In addition, to effectively learn vessel representation, we employ contrastive learning, namely contrastive-DARL (C-DARL), for further improvement \citep{wu2018unsupervised, chen2020simple, he2020momentum}. While DARL is trained to generate vessel masks via adversarial learning using the fractal masks, the input fractal masks and the real blood vessels have intrinsically different shapes and sizes. Accordingly, the contrastive loss function is designed to dissociate the estimated masks of real vessel images and the fractal masks while maximizing the similarity between the estimated masks of fractal-based synthetic vessel images and the fractal. This is achieved by leveraging contrastive unpaired translation (CUT) \citep{park2020contrastive} that computes the similarity of source and target in a patch-wise manner. 

Thanks to this simplification of the data preparation without requiring background images, our framework can be trained using multiple domains of two-dimensional (2D) blood vessel representations, such as X-ray angiography or retinal imaging. Experimental results show that our C-DARL model outperforms the comparative methods in vessel segmentation of various datasets. Also, when comparing our model to the DARL, our method achieves consistent improvement both on internal and external test data with respect to the training data. As the C-DARL provides vessel segmentation maps in real-time (0.176 seconds per frame), this holds great promise as a platform in clinical practices, upon validation of the integrity of the resulting representations. 
%%%%%% Contributions
In summary, the contributions of this paper are as follows:
\begin{itemize}
\item We introduce a label-free vessel segmentation method that can leverage multi-domain blood vessel images without requiring background images and provide vessel masks for those various datasets.
% We introduce a label-free vessel segmentation method that does not require the background images so that it can leverage multi-domain blood vessel images and provide vessel masks for those various datasets.
\item In contrast to the DARL, our proposed model applies contrastive learning in generating vessel segmentation maps, allowing the network to intensively learn vessel representations.
\item Extensive experimental results demonstrate that the proposed C-DARL is robust across diverse blood vessel data in a variety of clinical applications, and has superior performance with more efficiency than %to 
the existing self-supervised methods.
\end{itemize}

%%%%%%%%%%%%%%%%%%%%%%%%%%%%%%%%%%%%%%%%%%%%%%%%%%%%%%%%%%%%%%%%%%%%%%%%%%%%
%%%%%%%%%%%%%%%%%%%%%%%%%%%%%%%%%%%%%%%%%%%%%%%%%%%%%%%%%%%%%%%%%%%%%%%%%%%%
\section{Related works}

\subsection{Diffusion model}
The DDPM \citep{ho2020denoising} generates images by converting the Gaussian noise distribution into the data distribution through the Markov chain process. Specifically, for the forward diffusion, the data is corrupted by the noise as follows:
\begin{align}
    q(x_t|x_{t-1}) = \mathcal{N}(x_t;\sqrt{1-\beta_t}x_{t-1}, \beta_t I),
\end{align}
where $\beta_t$ is a scalar variance in the range of $[0,1]$. Then, through Markov chain, a noisy image $x_t$ for the data $x_0$ can be computed by:
\begin{align}
\label{eq:diff_forward}
    q(x_t|x_0) = \mathcal{N}(x_t;\sqrt{\alpha_t}x_{0}, (1-\alpha_t) I),
\end{align}
where $\alpha_t=\Pi_{s=1}^t(1-\beta_s)$.
For this corrupted image, the DDPM learns the reverse diffusion:
\begin{align}
    p_\theta (x_{t-1}|x_t) = \mathcal{N}(x_{t-1}; \bm\mu_\theta (x_t, t), \sigma_t^2 I),
\end{align}
where $\sigma_t$ is a scalar variance and $\bm\mu_\theta$ is a learnt mean computed by the network $G_\epsilon$:
\begin{align}
    \bm\mu_\theta (x_t, t) = \frac{1}{\sqrt{1-\beta_t}}\left(x_t-\frac{\beta_t}{\sqrt{1-\alpha_t}}G_\epsilon(x_t,t) \right).
\end{align}
Accordingly, one can obtain images from the Gaussian noise using the DDPM through the reverse diffusion process as follows: 
\begin{align}
x_{t-1}=\bm\mu_\theta (x_t, t)+\sigma_t z, 
\end{align}
where $z\sim\mathcal{N}(0, I)$.

This DDPM has been successfully adapted to various computer vision tasks including semantic image synthesis \citep{wang2022semantic, Huang_2023_CVPR}. Moreover, the potentials of learned representations from DDPM have been revealed through semantic segmentation, which effectively captures semantic features to improve segmentation performance \citep{baranchuk2021label, Rahman_2023_CVPR, asiedu2022decoder}.
% In this work, we leverage the potentials of DDPM for both the semantic image synthesis and the semantic segmentation tasks and achieve improved generalization performance on medical vessel segmentation.

\subsection{Self-supervised vessel segmentation using DARL} 
Semantic segmentation problems have been traditionally addressed using supervised learning.  Unfortunately, supervised learning performance is largely dependent on a huge amount of labels and associated costs, and limited and skilled resources \citep{fan2019accurate, yang2019deep}. Recently, self-supervised learning (SSL) methods have been actively investigated for mitigating this issue \citep{oquab2023dinov2, melas2022deep}. Nonetheless, a naive application of SSL is challenging to accurately extract vessel structure in various blood vessel images, which contain tiny vessel branches within highly interfering background signals. 

To address this, SSL methods tailored for the vessel segmentation task have been recently developed. In particular, \cite{kim2022diffusion} proposes a diffusion-based adversarial vessel segmentation method (DARL) that learns the background signal using the diffusion module, which effectively improves vessel segmentation performance with noise robustness. 

Specifically, the DARL model estimates vessel segmentation maps through the guidance of semantic image synthesis that incorporates the given pre-contrast background image and fractal vessel masks, as shown in Fig.~\ref{fig:method_motiv}(a). This is motivated by \cite{ma2021self} which synthesizes vessel images by adding the fractal masks to the background images and lets the network estimate vessel masks using the information of fractal-based synthetic vessel images. However, the DARL model has limitations in that 1) it still utilizes the background images as input, and 2) the vessel segmentation is learned through the adversarial loss by regarding the fractal masks as real and the network output as fake even though the fractals are different from the ground-truth vessel masks.

%%%%%%%%%%%%%%%%%%%%%%%%%%%%%%%%%%%%%%%%%%%%%%%%%%%%%%%%%%%%%%%%%%%%%%%%%%%%
%%%%%%%%%%%%%%%%%%%%%%%%%%%%%%%%%%%%%%%%%%%%%%%%%%%%%%%%%%%%%%%%%%%%%%%%%%%%
\section{Method}
\subsection{Motivation}
{To deal with the aforementioned issues, firstly, we propose a model that eliminates the need for background images, alleviating the constraint of using the angiography dataset, as shown in Fig.~\ref{fig:method_motiv}(b). This is based on the empirical observation that the diffusion module in DARL is effective in learning the sparsity of blood vessel structure but also in estimating the noise that captures the information of the given data distribution and enables diverse image synthesis.
Here, the diffusion model can estimate the noise by nulling out the learned image distribution regardless of the existence of vessels in the training data. Accordingly, as long as the vessel structures are sparsely distributed, the vessels can be regarded as outliers and represented in the diffusion module output when generating vessel masks. Therefore, the diffusion module can be trained using vessel images in various domains where vessel-free backgrounds are difficult to obtain.

\begin{figure}[!t]
\centering
\includegraphics[width=\linewidth]{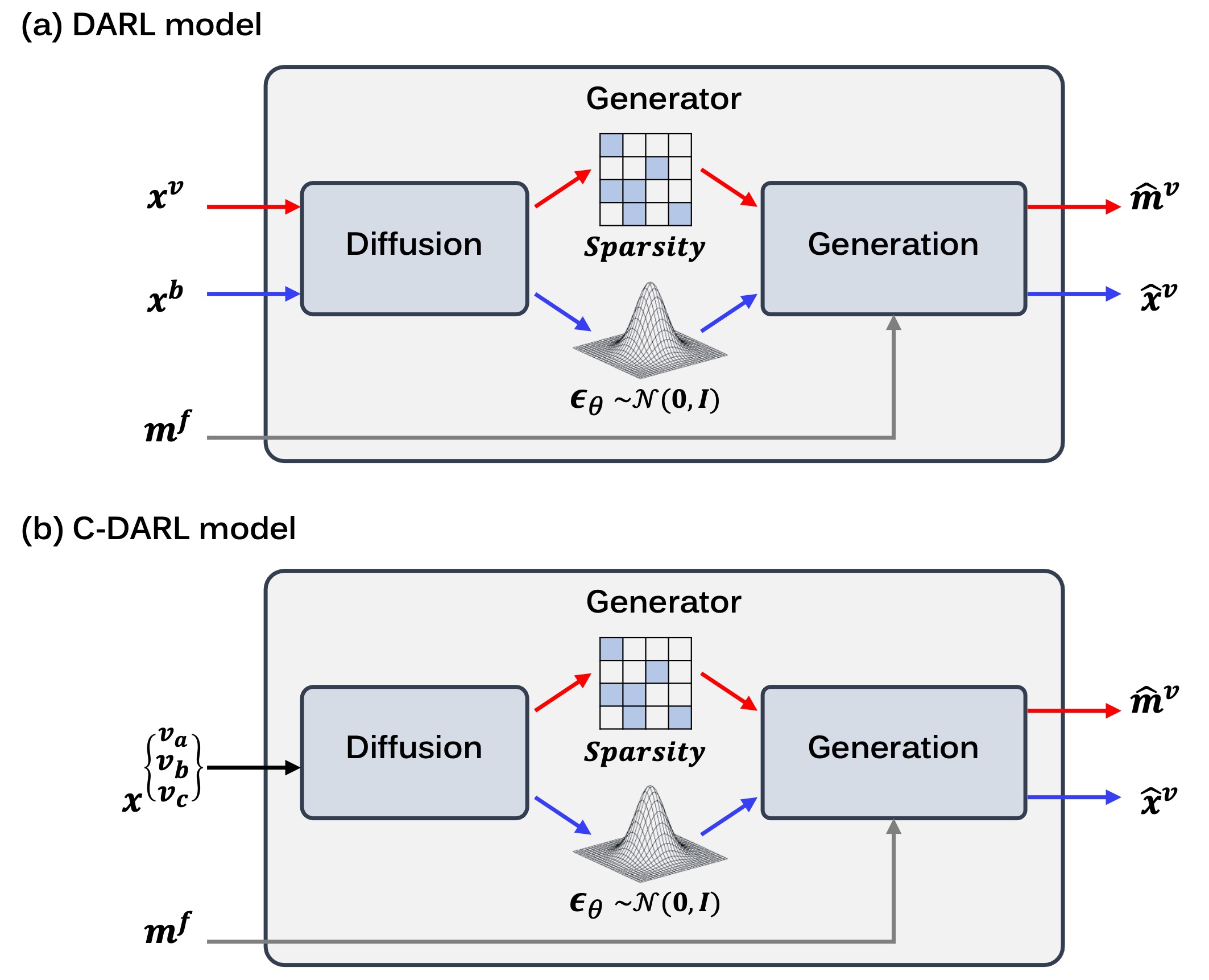}
\caption{Training pipelines of (a) the DARL model \citep{kim2022diffusion} and (b) our proposed C-DARL model. $x^v$ and $x^b$ are a real vessel image and a background image, respectively, and $m^f$ denotes a fractal mask. $\hat{m}^v$ is an estimated vessel segmentation mask, and $\hat{x}^v$ is a synthetic vessel image. For the C-DARL, the real vessel image in one of the various domains (e.g. $\{v_a, v_b, v_c\}$) can be fed into the model.}
\label{fig:method_motiv}
\end{figure}

\begin{figure*}[!t]
\centering
\includegraphics[width=0.9\linewidth]{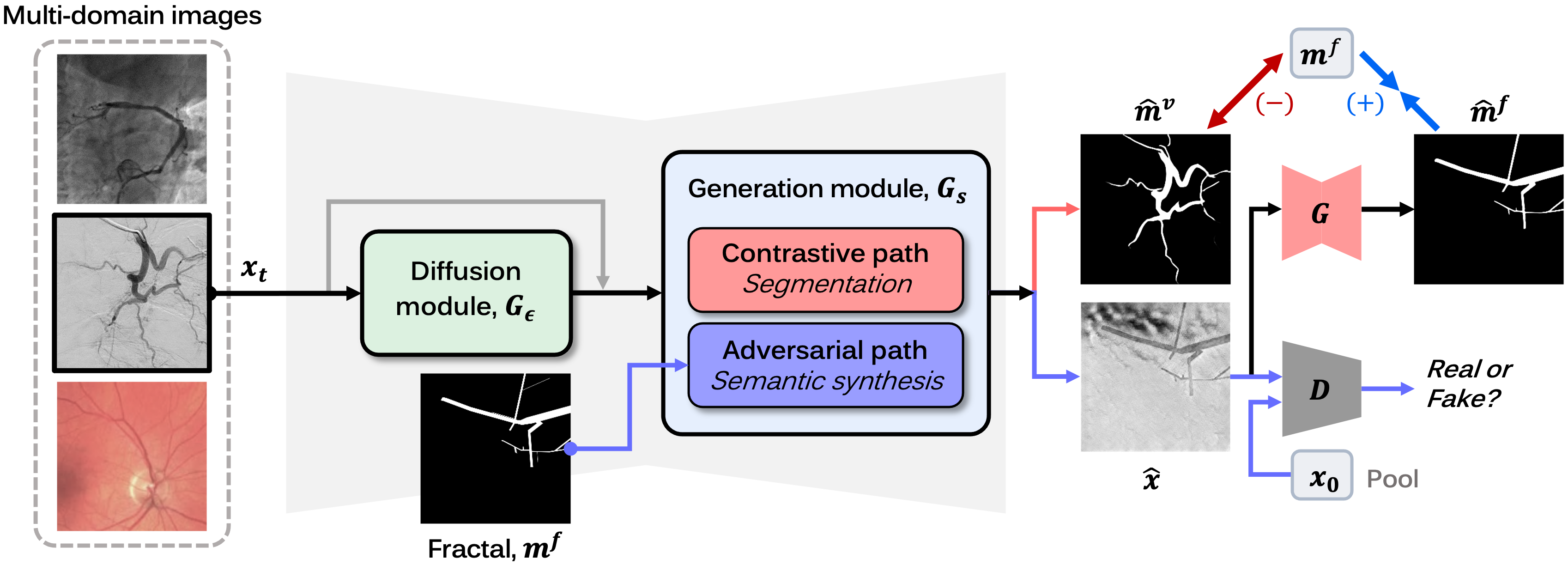}
\caption{Overall training framework of the proposed contrastive diffusion adversarial representation learning model (C-DARL).}
\label{fig:method}
\end{figure*}

Moreover, to further refine the segmentation accuracy, in the vessel segmentation path, we replace an adversarial loss of DARL with a contrastive loss \citep{chen2020simple, Zhong_2021_ICCV, Hu_2021_ICCV, Oh_2022_ACCV}. Specifically, by reflecting the fact that the fractal masks and real vessel masks have different features, we present a mask-based contrastive loss that utilizes the fractals and the estimated vessel masks as negative pairs, while using the fractals and the cyclically estimated segmentation masks as positive pairs. In particular, inspired by contrastive unsupervised translation (CUT) \citep{park2020contrastive}, which maximizes the mutual information of the patches at the same locations in the source and the target images, we employ the CUT-loss to maximize the structural patch similarity between the fractal mask (as a query) and the cyclically synthesized vessel mask (as a positive), and to disassociate the query signals from the segmented mask of the real vessel image (as a negative), allowing the model to learn vessel structure more effectively with no use of any labeled dataset.

%%%%%%%%%%%%%%%%%%%%%%%%%%%%%%%%%%%%%%%%%%%%%%%%%%%%%%%%%%%%%%%%%%%%%%%%%%%%
\subsection{Overall architecture}
The overall learning flow of C-DARL is illustrated in Fig.~\ref{fig:method}. The C-DARL model has a generator $G$ consisting of a diffusion module $G_\epsilon$ and a generation module $G_s$. When the diffusion module based on the DDPM estimates a latent feature by learning data distribution for various noisy levels, the generation module generates a vessel mask $\hat{m}^v$ and a synthetic vessel image $\hat{x}$. Here, the vessel image is generated using the latent of the diffusion module and the fractal mask $m^f$.

Then, in the path of vessel segmentation, contrastive learning is applied by using the fractal mask $m^f$ and the estimated vessel masks $(\hat{m}^v, \hat{m}^f)$, where $\hat{m}^f$ is generated by feeding the synthetic vessel image $\hat{x}$ into the generator through the cycle pathway. On the other hand, in the path of semantic image synthesis, the synthetic vessel image $\hat{x}$ and real vessel image $x$ are fed into a discriminator $D$ for adversarial learning. As described in Fig.~\ref{fig:method}, we denote the segmentation path as a \textit{contrastive} path and the image synthesis path as an \textit{adversarial} path. In the following, we describe the proposed method in detail.

%%%%%%%%%%%%%%%%%%%%%%%%%%%%%%%%%%%%%%%%%%%%%%%%%%%%%%%%%%%%%%%%%%%%%%%%%%%%
\subsubsection{Model input}
Let $\mathbf{X}=\bigcup_{k=1}^{k=K}X^k$ be a group of given blood vessel image datasets with $K$ different domains where $X^k$ has one or more images of the $k$-th domain, i.e. $X^k=\{x^{k_1}, x^{k_2}, ..., x^{k_N} \}$ with $k_N\ge 1$. For each iteration, our model randomly samples one of the multi-domain images $x^{k_n}_0$ among the given datasets. Then, for the model input, the image is corrupted by the forward diffusion (\ref{eq:diff_forward}):
\begin{align}\label{eq:input}
    x_t = \sqrt{\alpha_t}x^{k_n}_0 + \sqrt{1-\alpha_t}\epsilon,
\end{align}
where $t$ is the noisy level in range of $[0, T]$ and $\epsilon\sim\mathcal{N}(0, I)$. 
Using this perturbed image, the diffusion module $G_\epsilon$ learns the vessel image distribution and provides latent information to the generation module $G_s$.

%%%%%%%%%%%%%%%%%%%%%%%%%%%%%%%%%%%%%%%%%%%%%%%%%%%%%%%%%%%%%%%%%%%%%%%%%%%%
\subsubsection{Switchable SPADE layers}
When the diffusion module estimates the latent features, given the latent features concatenated with the perturbed image $x_t $ in channel dimension, the generation module generates vessel segmentation masks and synthetic vessel images simultaneously. This can be performed by the switchable SPADE (S-SPADE) layers proposed in the DARL model. Specifically, as shown in Fig.~\ref{fig:method_gen}, the generation module consists of $N$ residual blocks with S-SPADE layers that perform different normalization operations depending on whether or not the semantic layout is given to the model. For the feature maps $e$ from a shared encoder in the generation module, the residual blocks synthesize semantic images through the spatially adaptive normalization (SPADE) \citep{park2019semantic} if the semantic mask $m$ is given, whereas they generate segmentation masks through the instance normalization (InsNorm) \citep{ulyanov2016instance} otherwise:
 \begin{equation}
    e=
    \begin{cases}
      \text{SPADE}(e, m), & \text{if semantic mask $m$ is given,}  \\
      \text{InsNorm}(e), & \text{otherwise}.
    \end{cases}
  \end{equation}

\begin{figure}[t!]
\centering
\includegraphics[width=\linewidth]{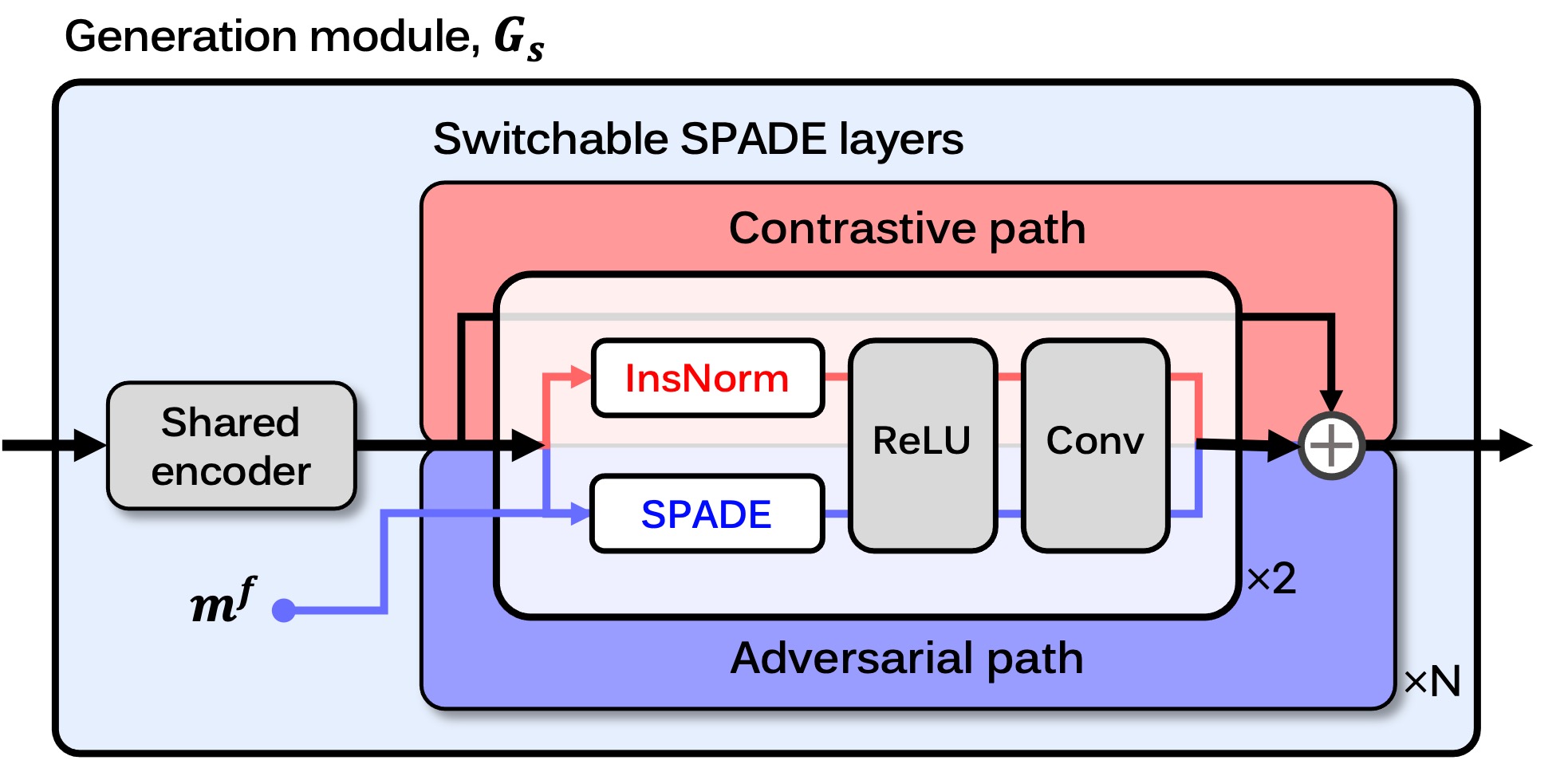}
\caption{Detailed architecture of the generation module $G_s$ that has a shared encoder and $N$ residual blocks composed of switchable SPADE layers, ReLU, and convolution (Conv) layers. If a semantic layout $m^f$ is given to the generation module, the SPADE layer is activated. Otherwise, the instance normalization (InsNorm) is activated.}
\label{fig:method_gen}
\end{figure}

Thus, the contrastive path activates the InsNorm and estimates the vessel masks, and at the same time, the adversarial path, which takes the fractal masks $m^f$, activates the SPADE layer and generates the synthetic images. Also, by sharing all network parameters except for the S-SPADE layers in the two paths of the generation module, our model can synergistically learn vessel representation through semantic image synthesis. This also ensures that the estimated vessel structure from the contrastive path becomes a semantically meaningful negative pair with respect to the synthetic fractal mask.

Therefore, although the contrastive path does not need to learn the reverse process from the pure Gaussian noise in that this path estimates the vessel masks of the input image, the corrupted input image is given also to the contrastive path to share the same generation module with SPADE layer. Instead, by setting the maximum noisy level $T$ to be smaller than that for the adversarial path, we design the model to estimate vessel masks for noisy input images.

%%%%%%%%%%%%%%%%%%%%%%%%%%%%%%%%%%%%%%%%%%%%%%%%%%%%%%%%%%%%%%%%%%%%%%%%%%%%
\subsection{Loss formulation}

\begin{figure}[!ht]
\centering
\includegraphics[width=\linewidth]{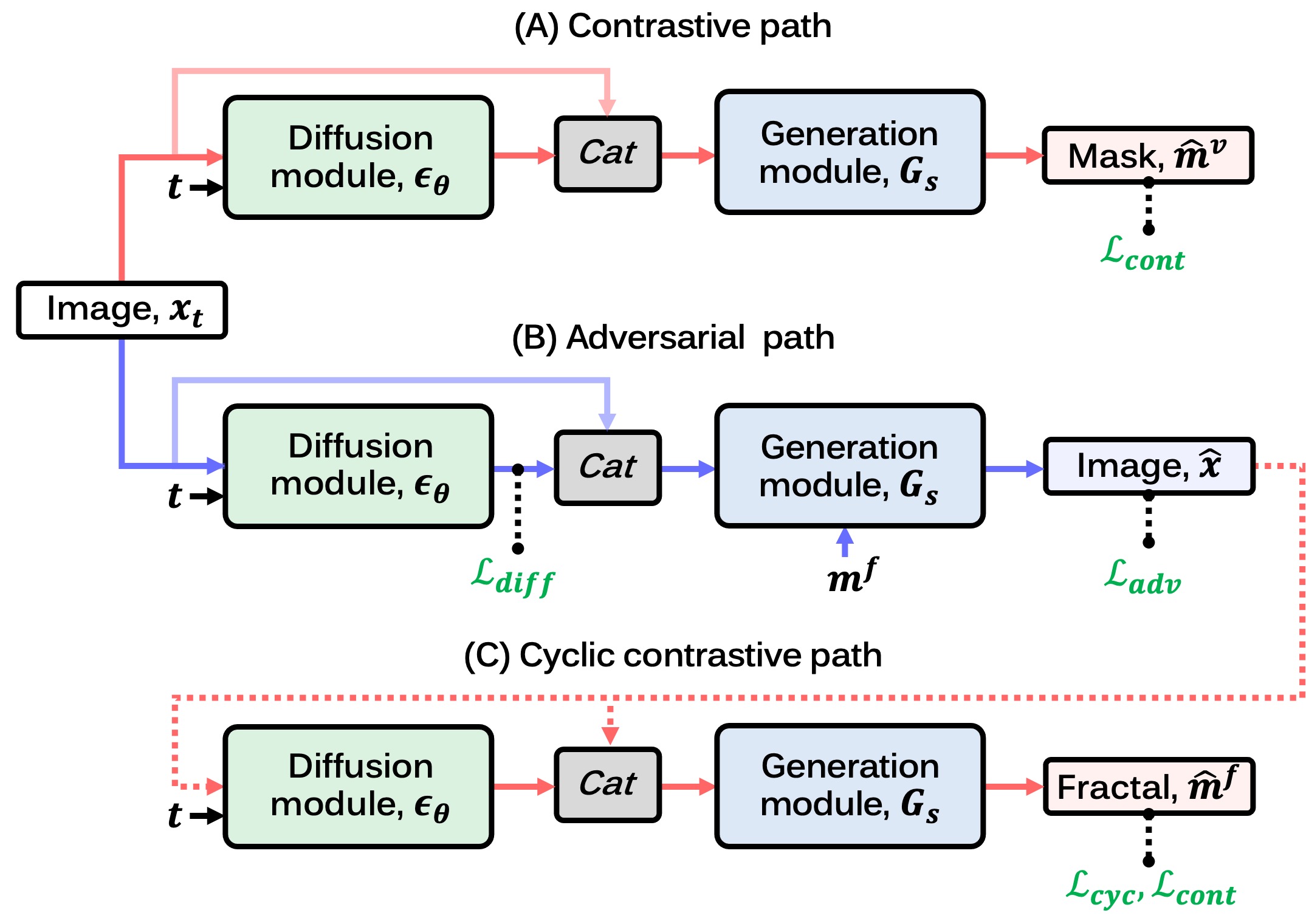}
\caption{Diagram of loss formulation to train our proposed C-DARL. \textit{Cat} denotes the channel-wise concatenation of two inputs. }
\label{fig:loss}
\end{figure}
Fig.~\ref{fig:loss} shows our loss function with training flow. The proposed C-DARL model has three paths in the training phase:
% To train the proposed C-DARL model, we employ contrastive learning and adversarial learning frameworks. Specifically, as shown in Fig.~\ref{fig:loss}, the training flow of our model has three paths: 
(A) the contrastive path estimates the vessel segmentation mask $\hat{m}^v$ that is used to compute a mask-based contrastive loss $\mathcal{L}_{cont}$, (B) the adversarial path learns data distribution through a diffusion loss $\mathcal{L}_{diff}$ and generates the vessel image $\hat{x}$ based on the fractal layout $m^f$ via an adversarial loss $\mathcal{L}_{adv}$, and (C) the cyclic contrastive path feeds the fractal-based synthetic image into the model and segments the fractal region $\hat{m}^f$ that is utilized in a cycle loss $\mathcal{L}_{cyc}$ and the contrastive loss $\mathcal{L}_{cont}$. A detailed description of each loss function is as follows. 

%%%%%%%%%%%%%%%%%%%%%%%%%%%%%%%%%%%%%
\subsubsection{Diffusion loss}
The diffusion loss aims to learn input data distribution through the diffusion module $G_\epsilon$, yielding latent features including meaningful information about the input. For a given vessel image $x_0$, a noise $\epsilon\sim\mathcal{N}(0, I)$, and a time step $t$ uniformly sampled from $[0, T]$, by following the DDPM training scheme \citep{ho2020denoising}, the loss can be represented as:
\begin{align}
    \mathcal{L}_{diff}=\mathbb{E}_{x_0, \epsilon, t} \left[||G_\epsilon(\sqrt{\alpha_t}x_0 + \sqrt{(1-\alpha_t)}\epsilon, t) - \epsilon ||^2 \right].
\end{align}

As aforementioned, since the input image is perturbed within the full range of noisy levels in the adversarial path compared to the contrastive path, the diffusion loss is calculated only in the adversarial path.% This allows the diffusion module to efficiently learn the distribution of vessel data.

%%%%%%%%%%%%%%%%%%%%%%%%%%%%%%%%%%%%%
\subsubsection{Mask-based contrastive loss}
To address the limitation of not having real vessel masks in our label-free training scheme, we employ a patch-based contrastive learning objective \citep{park2020contrastive} that maximizes the mutual information of corresponding patches between two images, which utilizes a noise contrastive estimation method \citep{oord2018representation}. Here, instead of using the image features from the network encoder, we design a \textit{mask-based} contrastive loss that leverages the segmentation masks. 

Specifically, based on the observation that the real blood vessel and the fractal masks have different features of shapes and sizes, for the fractal $m^f$ as a query, we refer to the estimated vessel mask $\hat{m}^v$ in the contrastive path as negatives (See Fig.~\ref{fig:method}). In contrast, since the cyclic contrastive path estimates the fractal embedded in the synthetic vessel image, we regard the segmented fractal mask $\hat{m}^f$ as a positive of the query. Then, the model is trained for corresponding patches of $m^f$ and $\hat{m}^f$ to be more strongly associated than those of $m^f$ and $\hat{m}^v$ using our contrastive loss.

More specifically, for each segmentation mask $m\in\mathbb{R}^{1\times HW}$, we obtain $R$ different-sized tensors, in which each tensor is obtained by folding the mask as $m_{r}\in\mathbb{R}^{{p_r}^2 \times \frac{H}{p_r}\frac{W}{p_r}}$ where $p_r$ is a division factor for $r\in\{1,2, ..., R\}$. Then, the tensor is fed into a light-weight network $H_r$ composed of two fully-connected layers, generating a stack of the features $\{h_r(m)\}_R=\{H_1(m_1), H_2(m_2), ..., H_R(m_R)\}$ where $h_r(m)\in\mathbb{R}^{C_r \times \frac{H}{p_r}\frac{W}{p_r}}$ is the feature with $C_r$ channels for the $r$-th tensor. Through this process, we can get the feature stacks of $\{h_r(m^f)\}_R$, $\{h_r(\hat{m}^f)\}_R$, and $\{h_r(\hat{m}^v)\}_R$ for the masks of $m^f$, $\hat{m}^f$, and $\hat{m}^f$, respectively. By randomly selecting $Q_r$ spatial locations in the range of $[0, \frac{H}{p_r}\frac{W}{p_r}]$, the contrastive loss is computed by:
\begin{align}
    \mathcal{L}_{cont}=\mathbb{E}_{m^f, \hat{m}^f, \hat{m}^v}\sum_{r=1}^R\sum_{q=1}^{Q_r}\ell_{MI}\left(h^q_{r}(m^f), h^q_{r}(\hat{m}^f), h^q_{r}(\hat{m}^v)\right),
\end{align}
where $\ell_{MI}$ is the mutual information using the cross-entropy loss as:
\begin{align}
    \ell_{MI}(u, u^+, u^-)=-log\left[\frac{\text{exp}(u\cdot u^+/\tau)}{\text{exp}(u\cdot u^+/\tau)+\sum_i \text{exp}(u\cdot u_i^-/\tau)} \right],
\end{align}
where $\tau$ is a temperature scaling the distances between the query and other positives/negatives. This calculates the probability that the positive patches are selected over the negative patches at specific locations, enabling the model to generate vessel masks not very similar to the fractal masks while learning the mask features. In our experiments, we set $R=3$, and $(p_1, p_2, p_3)=(4,8,16)$.

\begin{figure*}[t!]
\centering
\includegraphics[width=\linewidth]{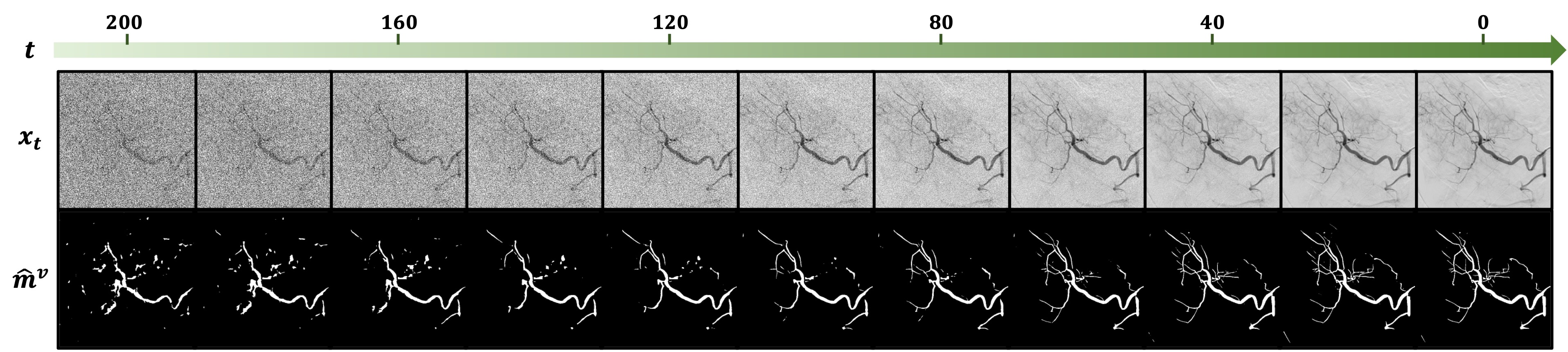}
\caption{Inference of the blood vessel segmentation according to the noisy level $t$. For given perturbed input images $x_t$, the proposed C-DARL model generates vessel masks $\hat{m}^v$ in a single step. We show an example of the vessel segmentation of the hepatic angiogram.}
\label{fig:infer}
\end{figure*}

%%%%%%%%%%%%%%%%%%%%%%%%%%%%%%%%%%%%%
\subsubsection{Adversarial loss}
In the adversarial path, our model generates the synthetic vessel image using the semantic fractal mask and the noisy features generated by the diffusion module. Since real vessel images exist in the training phase, we apply adversarial learning to the model and enable the generator $G$ to output realistic images while fooling the discriminator $D$ that distinguishes real and fake images. By employing LSGAN \citep{mao2017least}, the adversarial loss of the generator can be represented as:
\begin{align}
    \mathcal{L}_{adv}^G = \mathbb{E}_{x_t,m^f}\left[(D(G(x_t,m^f))-1)^2\right],
\end{align}
and the adversarial loss of the discriminator can be written as:
\begin{align}
    \mathcal{L}_{adv}^D = \frac{1}{2}\mathbb{E}_{x_0}\left[(D(x_0)-1)^2\right]+\frac{1}{2}\mathbb{E}_{x_t,m^f}\left[(D(G(x_t,m^f))^2\right].
\end{align}
Through this loss function, the model is trained to generate the synthetic vessel image $\hat{x}$ that is indistinguishable from real data $x$ by the discriminator.

%%%%%%%%%%%%%%%%%%%%%%%%%%%%%%%%%%%%%
\subsubsection{Cycle loss}
Recall that when the fractal-based synthetic vessel image is generated, in the cyclic contrastive path, the proposed model takes this synthetic image and generates the fractal segmentation mask $\hat{m}^f$. Here, in addition to using the mask $\hat{m}^f$ in the contrastive loss, to learn semantic information about blood vessels in the vessel medical data, we utilize the estimated fractal mask in the cycle loss that computes a distance between the real and fake images. As we handle the mask as an image, the cycle loss can be formulated using $l_1$ norm:
\begin{align}
    \mathcal{L}_{cyc} = \mathbb{E}_{x_t,m^f}\left[||G(G(x_t,m^f),\text{0})-m^f||_1 \right].
\end{align}
Accordingly, the model can capture vessel information and estimate the mask for vessel-like regions even though there is no ground-truth vessel mask of real data.

%%%%%%%%%%%%%%%%%%%%%%%%%%%%%%%%%%%%%
\subsubsection{Full loss formulation}
Using the diffusion loss, the contrastive loss, the adversarial loss, and the cycle loss, our C-DARL model is trained in an end-to-end learning manner. Hence, the complete loss function of the generator can be defined by:
\begin{align}
\label{eq:loss}
    loss = \mathcal{L}_{diff} + \lambda_\alpha\mathcal{L}_{cont} + \lambda_\beta\mathcal{L}_{adv}^G + \lambda_\gamma\mathcal{L}_{cyc},
\end{align}
where $\lambda_\alpha$, $\lambda_\beta$, and $\lambda_\gamma$ are hyper-parameters that control each loss function.

%%%%%%%%%%%%%%%%%%%%%%%%%%%%%%%%%%%%%%%%%%%%%%%%%%%%%%%%%%%%%%%%%%%%%%%%%%%%
\subsection{Inference}
Compared to conventional diffusion models, which generate images from pure Gaussian noise through the iterative reverse process, the DARL model provides a segmentation mask in one step in that the model is trained to estimate the mask for a given input image. Similarly, the proposed C-DARL model also generates the mask for a given vessel medical image in a single step. 
In other words, for a perturbed vessel image $x_t$, our model outputs the vessel segmentation mask $\hat{m}^v$ through the contrastive path. Here, while the model can estimate the mask for any corrupted image with a noise level $t\in[0, T]$ as shown in Fig.~\ref{fig:infer}, we evaluate our method using the clean image $x_0$ which can be considered as a target image for the diffusion model. We further discuss the segmentation performance according to the noise level in {Section~\ref{sec:dis_nlevel}}.

%%%%%%%%%%%%%%%%%%%%%%%%%%%%%%%%%%%%%%%%%%%%%%%%%%%%%%%%%%%%%%%%%%%%%%%%%%%%
%%%%%%%%%%%%%%%%%%%%%%%%%%%%%%%%%%%%%%%%%%%%%%%%%%%%%%%%%%%%%%%%%%%%%%%%%%%%
\section{Experiments and results}

\begin{table}[!t]
    \caption{The number of data used in each training, validation (val), and test phase.}
    \centering
    \resizebox{\linewidth}{!}{
    \begin{tabular}{wl{0.6cm}wl{2.1cm}wc{0.9cm}wc{1.0cm}wc{1.0cm}wc{0.8cm}wc{0.5cm}wc{0.3cm}wc{0.95cm}wc{0.95cm}}%{lcccccccc}
        \toprule
        & \multirow{2}{*}{\bf{Input}} & \multicolumn{3}{c}{\bf{CA}} & \multicolumn{1}{c}{\bf{AA}} & \multicolumn{4}{c}{\bf{RI}}  \\ 
         \cmidrule(r){3-5} \cmidrule(r){6-6}\cmidrule(r){7-10}
        & & XCAD & 134XCA & 30XCA & {AAIR} & UWF & FP & DRIVE & STARE \\ 
        \midrule
        \multirow{2}{*}{\bf{Train}} & Image $\x$ & 1,621 & - & - & 327 & 451 & 745 & - & - \\
        % & Mask; $\m^{GT}$ & 0 & 0 & 0 & - & - & 0 & - & - \\
        & Fractal mask $\m^f$ & 1,621 & - & - & - & - & - & - \\
        \midrule
        \multirow{2}{*}{\bf{Val}} & Image $\x$ & 12 & - & - & - & - & - & - & - \\
        & Vessel label $\m^{v}$ & 12 & - & - & - & - & - & - & - \\
        \midrule
        \multirow{2}{*}{\bf{Test}}& Image $\x$ & 114 & 134 & 30 & {-} & - & - & 20 & 20 \\
        & Vessel label $\m^{v}$ & 114 & 134 & 30 & {-} & - & - & 20 & 20 \\
        \bottomrule
        \multicolumn{10}{l}{CA: Coronary angiography, \quad AA: Abdomen angiography, \quad RI: Retinal imaging}
    \end{tabular}
    }
    \label{tab:dataset}
\end{table}

%%%%%%%%%%%%%%%%%%%%%%%%%%%%%%%%%%%%%%%%%%%%%%%%%%%%%%%%%%%%%%%%%%%%%%%%%%%%
\subsection{Experimental setup}

\subsubsection{Datasets}
To train our C-DARL model for blood vessel segmentation, we utilize a variety of vessel images from different medical domains, including coronary arteriograms, abdominal pancreatic and hepatic arteriograms, and retinal fundus photography images. Also, we test the proposed model using several benchmark datasets of blood vessel segmentation. The number of data used in the training, validation, and test phases is summarized in Table~\ref{tab:dataset}, and detailed descriptions and preprocessing methods of each dataset are as follows.

\paragraph{XCAD dataset}
The X-ray angiography coronary artery disease (XCAD) dataset \citep{ma2021self} contains 1,621 coronary angiography images taken during stent placement. Also, using the fractal module, we synthesize 1,621 fractal masks for model training. For an additional 126 angiograms and the corresponding vessel segmentation masks labeled by experienced radiologists, we use 12 pairs of images and masks for the validation set, and the remaining 114 pairs for the test set. Each image has $512\times 512$ sized resolution, and we subsample it into $256\times 256$. In addition, by following the fractal synthesis method in \cite{ma2021self}, we generate 1,621 synthetic fractal masks with $512\times 512$ size, in which the fractals have various shapes and thicknesses.

\paragraph{134XCA dataset}
To test the model for vessel segmentation of coronary angiography images using externally distributed data over the training dataset, we use the 134 XCA dataset \citep{app9245507} that provides 134 coronary angiograms with ground-truth vessel masks, where the masks are obtained by an expert cardiologist. We resize the images to $512\times 512$ and extract vessel masks using 4 subsampled $256\times 256$ images. The full vessel masks are then obtained by un-subsampling the 4 masks into one.

\paragraph{30XCA dataset}
We also utilize the 30XCA dataset \citep{HAO2020172} in the evaluation of coronary angiography segmentation. This dataset has 30 X-ray coronary angiography series, and one image for each series and its corresponding vessel segmentation masks annotated by experts are used for the test. The images are processed as same as the 134XCA dataset.

\paragraph{AAIR dataset}
As one of the multi-domain angiography images, we use 327 abdominal angiography images that were obtained from 42 subjects at the National Institutes of Health Clinical Center. The angiography series were obtained with 2 frames per second (fps) during arteriography, embolization or calcium stimulation and show arteries in the abdomen such as the celiac, pancreatic, and hepatic vasculatures. For each scan, we select one frame in which the blood vessels were most visible and manually cropped the region that includes the vessels. We call these collected data the AAIR dataset. In the training phase, we randomly extract image patches with a size of $256\times 256$.
% [Chemo embolization] + [Embolization] + [Abdominal w/ calcium stimulation]

\paragraph{UWF and FP datasets}
In addition to the coronary and abdominal angiograms, our model is trained using retinal images of the ultra-widefield (UWF) data and the fundus photograph (FP) data which are provided by \cite{yoo2020deep}. The UWF dataset has 451 normal or pathologic retinal images with artifacts, and the FP dataset has 745 retinal images without artifacts. We use these retinal data for training the model. For data processing, we resize each image to $512\times 512$ and crop the center area to $256\times 256$. Also, we convert the RGB scale images to grayscale for the data to have one channel.

\paragraph{DRIVE dataset}
To evaluate the model for retinal imaging data, we use the DRIVE dataset \citep{staal2004ridge} that provides 20 retinal images and their vessel segmentation labels. We transform the images from RGB scale to grayscale and resize them to $768\times768$. Then we extract non-overlapping patches with a size of $256\times 256$ and obtain the vessel masks for nonzero areas of those patches.

\paragraph{STARE dataset}
For the retinal image domain, the model is also tested using the STARE dataset \citep{hoover2000locating}. This is composed of 20 retinal images and human-labeled blood vessel segmentation masks. As with the DRIVE dataset, the vessel masks are estimated for given $256\times 256$ patches from the rescaled grayscale retinal images.

%%%%%%%%%%%%%%%%%%%%%%%%%%%%%%%%%%%%%%%%%%%%%%%%%%%%%%%%%%%%%%%%%
\subsubsection{Implementation details}
Our proposed model is implemented using the PyTorch \citep{paszke2019pytorch} in Python. For the network architectures of the diffusion module and the generation module, we employ the DDPM \citep{ho2020denoising} and the SPADE \citep{park2019semantic}, respectively. The DDPM network has a U-Net-like structure and takes an embedding vector of the time as well as the noise-corrupted image. Also, the SPADE model has a shallow encoder and a decoder with SPADE blocks, in which the SPADE layer is replaced with the S-SPADE layer for our C-DARL model. Also, for the discriminator, we utilize PatchGAN \citep{isola2017image} to distinguish real and generated fake images in patch levels. 

We set the range of noisy time steps as [0, 2000] to linearly schedule the noise levels from $10^{-6}$ to $10^{-2}$. In the contrastive path, we set the maximum noisy level to 200, allowing the model to perform robust segmentation even on noisy images. The hyperparameters of our loss function are set as $\lambda_\alpha=4\times10^{-4}$, $\lambda_\beta=0.2$, and $\lambda_\gamma=2$. For the inputs of multi-domain images, as shown in Table~\ref{tab:dataset}, we use all images from the XCAD, AAIR, UWF, and FP datasets. The images are rescaled into [-1, 1] and augmented using random horizontal or vertical flips and rotation at 90 degrees. 

To optimize our model, we adopt the Adam algorithm \citep{kingma2014adam} with a learning rate of $1\times 10^{-5}$. The model is trained for 150 epochs using a single GPU card of Nvidia A100-SXM4-40GB. Then, we test the proposed method using the network weights that achieve the best performance on the validation dataset.

\subsubsection{Evaluation}
For baseline methods, we adopt several self-supervised approaches that can use the pseudo fractal masks for a fair comparison: Deep Adversarial (DA) \citep{mahmood2019deep}, Self-Supervised Vessel Segmentation (SSVS) \citep{ma2021self}, and Diffusion Adversarial Representation Learning (DARL) \citep{kim2022diffusion}. For the models of SSVS and DARL that require background images acquired before contrast agent injection, we use only coronary angiography (CA) data from the XCAD dataset. For the DA method, we train two models: one is to only use the CA data, and the other is to use the same amount of training data as our method. All these baseline methods are trained under identical experimental conditions to the proposed C-DARL.
The segmentation performance is evaluated quantitatively using several metrics: Intersection over Union (IoU), Dice similarity coefficient, and Precision. 
% We further compare our C-DARL to several unsupervised methods, which do not require any synthetic fractal masks, e.g., Spatial-Guided Clustering (SGC) \citep{ahn2021spatial} and Redrawing \citep{chen2019unsupervised}. All the unsupervised baseline methods are trained under identical experimental conditions to C-DARL, except for a condition of not utilizing synthetic fractal masks. 

\begin{table}[!t]
    \caption{Quantitative evaluation results of vessel segmentation performance for the comparison study with the self-supervised methods.} 
    \centering
    \resizebox{\linewidth}{!}{
    \begin{tabular}{llccccc}
    
     \toprule

        \multirow{2}{*}{\bf{Datset}} & \multirow{2}{*}{\bf{Metric}} & \multicolumn{3}{c}{\bf{w/ only CA data}}  & \multicolumn{2}{c}{\bf{w/ multi-domain data}} \\
        \cmidrule(l){3-5}  \cmidrule(l){6-7} 
        & & DA & SSVS & DARL & DA & C-DARL \\ 
        \midrule
        
        \multicolumn{7}{l}{Coronary angiography (CA)}
        \\ \midrule
        \multirow{3}{*}{XCAD}  &  IoU  &  $0.375_{\pm 0.066}$  &  $0.410_{\pm 0.087}$  &  $0.471_{\pm 0.076}$  &  $0.303_{\pm 0.085}$  &   {$\bf0.498_{\pm 0.086}$}  \\  
         &  Dice  &  $0.542_{\pm 0.073}$  &  $0.575_{\pm 0.091}$  &  $0.636_{\pm 0.072}$  &  $0.459_{\pm 0.101}$  &  {$\bf0.661_{\pm 0.079}$}  \\  
         &  Precision  &  $0.557_{\pm 0.115}$  &  $0.590_{\pm 0.119}$  &  $0.701_{\pm 0.115}$  &  $0.449_{\pm 0.119}$  &  {$\bf0.750_{\pm 0.108}$}  \\ 
         \cmidrule(l){1-1} \cmidrule(l){2-2} \cmidrule(l){3-5} \cmidrule(l){6-7}   

        \multirow{3}{*}{134 XCA}  &  IoU  &  {$0.257_{\pm 0.152}$}  &  {$0.384_{\pm 0.078}$}  &  {$0.509_{\pm 0.105}$}  &  $0.286_{\pm 0.200}$  &  {$\bf0.545_{\pm 0.094}$}  \\ 
         &  Dice  &  {$0.384_{\pm 0.206}$}  &  {$0.550_{\pm 0.082}$}  &  {$0.667_{\pm 0.098}$}  &  $0.407_{\pm 0.240}$  &  {$\bf0.700_{\pm 0.087}$}  \\  
         &  Precision  &  {$0.490_{\pm 0.187}$}  &  {$0.534_{\pm 0.094}$}  & {$0.645_{\pm 0.131}$}  &  $0.379_{\pm 0.225}$  &  {$\bf0.673_{\pm 0.128}$}  \\ 
         \cmidrule(l){1-1} \cmidrule(l){2-2} \cmidrule(l){3-5} \cmidrule(l){6-7}  
                
        \multirow{3}{*}{30 XCA}  &  IoU  &  $0.298_{\pm 0.109}$  &  $0.324_{\pm 0.146}$  &  $0.427_{\pm 0.184}$  &  $0.353_{\pm 0.083}$  &  {$\bf0.453_{\pm 0.061}$}  \\ 
         &  Dice  &  $0.447_{\pm 0.148}$  &  $0.468_{\pm 0.193}$  &  $0.572_{\pm 0.205}$  &  $0.516_{\pm 0.091}$  &  {$\bf0.621_{\pm 0.059}$}  \\  
         &  Precision  &  $0.612_{\pm 0.174}$  &  $0.613_{\pm 0.212}$  &  $0.729_{\pm 0.152}$  &  $0.659_{\pm 0.121}$  &  {$\bf0.826_{\pm 0.083}$}  \\ \midrule 
                
        \multicolumn{7}{l}{Retinal Imaging (RI)} \\ 
        \midrule
         \multirow{3}{*}{DRIVE}  &  IoU  &  {$0.180_{\pm 0.029}$}  &  {$0.229_{\pm 0.045}$}  &  {$0.313_{\pm 0.056}$}  &  $0.336_{\pm 0.040}$  &  {$\bf0.345_{\pm 0.061}$} \\ 
         &  Dice  &  {$0.304_{\pm 0.042}$}  &  {$0.371_{\pm 0.058}$}  &  {$0.474_{\pm 0.065}$}  &  $0.501_{\pm 0.045}$  &  {$\bf0.510_{\pm 0.067}$} \\ 
         &  Precision  &  {$0.472_{\pm 0.053}$}  &  {$0.632_{\pm 0.075}$}  & {$0.809_{\pm 0.099}$}  &  $0.687_{\pm 0.052}$  &  {$\bf0.904_{\pm 0.090}$} \\ 
         \cmidrule(l){1-1} \cmidrule(l){2-2} \cmidrule(l){3-5} \cmidrule(l){6-7}  
        
         \multirow{3}{*}{STARE}  &  IoU  &  {$0.217_{\pm 0.071}$}  &  {$0.293_{\pm 0.089}$}  &  {$\bf0.383_{\pm 0.136}$}  &  $0.328_{\pm 0.092}$  &  {$0.367_{\pm 0.133}$} \\  
           &  Dice  &  {$0.351_{\pm 0.099}$}  &  {$0.446_{\pm 0.111}$}  &  {$\bf0.539_{\pm 0.151}$} &  $0.486_{\pm 0.110}$  &  {$0.522_{\pm 0.154}$} \\  
           &  Precision  &  {$0.450_{\pm 0.111}$}  &  {$0.571_{\pm 0.114}$}  &  {$0.671_{\pm 0.192}$}  &  $0.585_{\pm 0.117}$  &  {$\bf0.736_{\pm 0.179}$} \\ 
        \bottomrule
              
    \end{tabular}
    }\label{tab:compare_self}
    \vspace{-0.2cm}
\end{table}

\begin{figure*}[!t]
\centering
\includegraphics[width=1\linewidth]{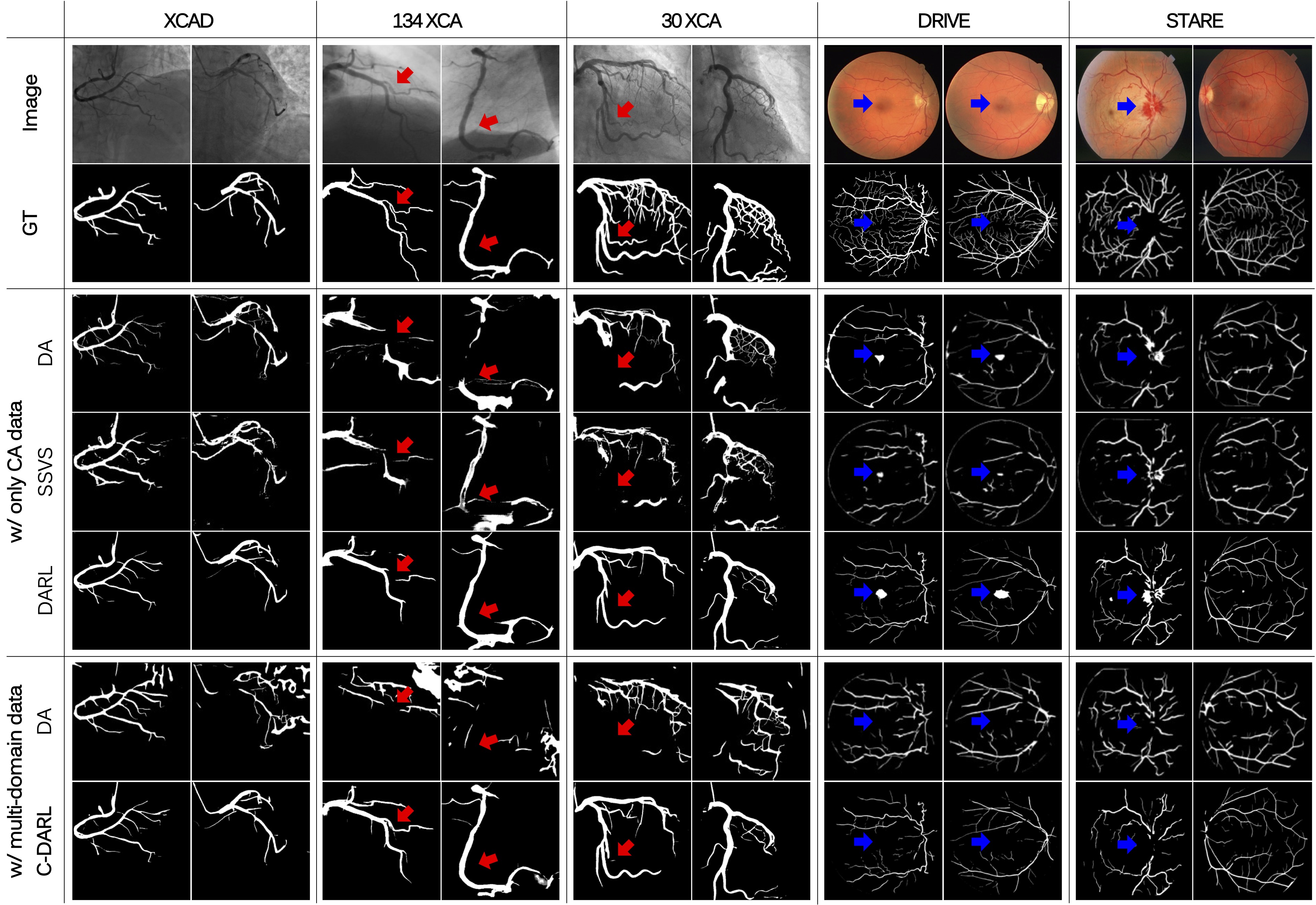}
\caption{Visual comparisons of the vessel segmentation results on various vessel datasets. Red and blue arrows indicate remarkable parts of tiny vessel structures and false positive artifacts, respectively.}
\label{fig:main}
\end{figure*}

%%%%%%%%%%%%%%%%%%%%%%%%%%%%%%%%%%%%%%%%%%%%%%%%%%%%%%%%%%%%%%%%%%%%%%%%%%%%
\subsection{Results} 
\subsubsection{Comparison study}
\label{sec:result_comp}
For the comparison study with the existing models, we evaluate the blood vessel segmentation performance on coronary angiography (CA) datasets and retinal imaging (RI) datasets. As reported in Table~\ref{tab:compare_self}, our model achieves state-of-the-art (SOTA) performance on most of the datasets compared to the baseline methods of self-supervised vessel segmentation. In particular, although the proposed C-DARL model is trained using various domains of vessel images in contrast to the previous model of DARL, our performance is higher than the DARL model. This result indicates that the proposed contrastive learning framework using multi-domain images improves not only the segmentation performance but also the generalization performance on unseen datasets. 

In Fig.~\ref{fig:main}, we can observe that the segmentation performance on tiny vessel regions is significantly improved (see red arrows), which demonstrates the advantages of our contrastive learning that effectively differentiates real vessel structure from the pseudo fractal mask signal. 
Moreover, our C-DARL mitigates false positive artifacts compared to the existing models trained with only the CA dataset (see blue arrows) and shows the most promising precision metrics. This shows the training efficiency of the proposed label-free framework which is capable of incorporating various data distributions with no use of paired background images.

\begin{figure*}[!t]
\centering
\includegraphics[width=\linewidth]{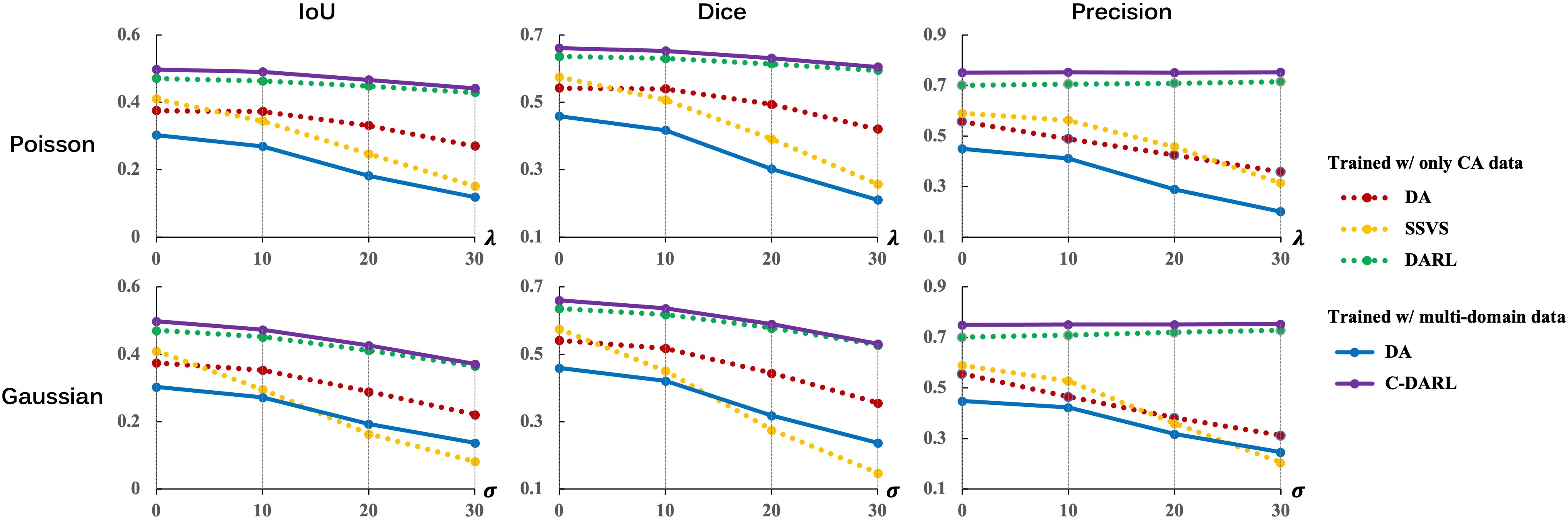}
\caption{Quantitative results of vessel segmentation on various noise-corrupted image scenarios. The top row shows the evaluation results according to $\lambda$ level for Poisson noisy images, and the bottom row shows the results according to $\sigma$ level for Gaussian noisy images.}
\label{fig:compare_noise}
\end{figure*}

\begin{figure}[!t]
\centering
\includegraphics[width=\linewidth]{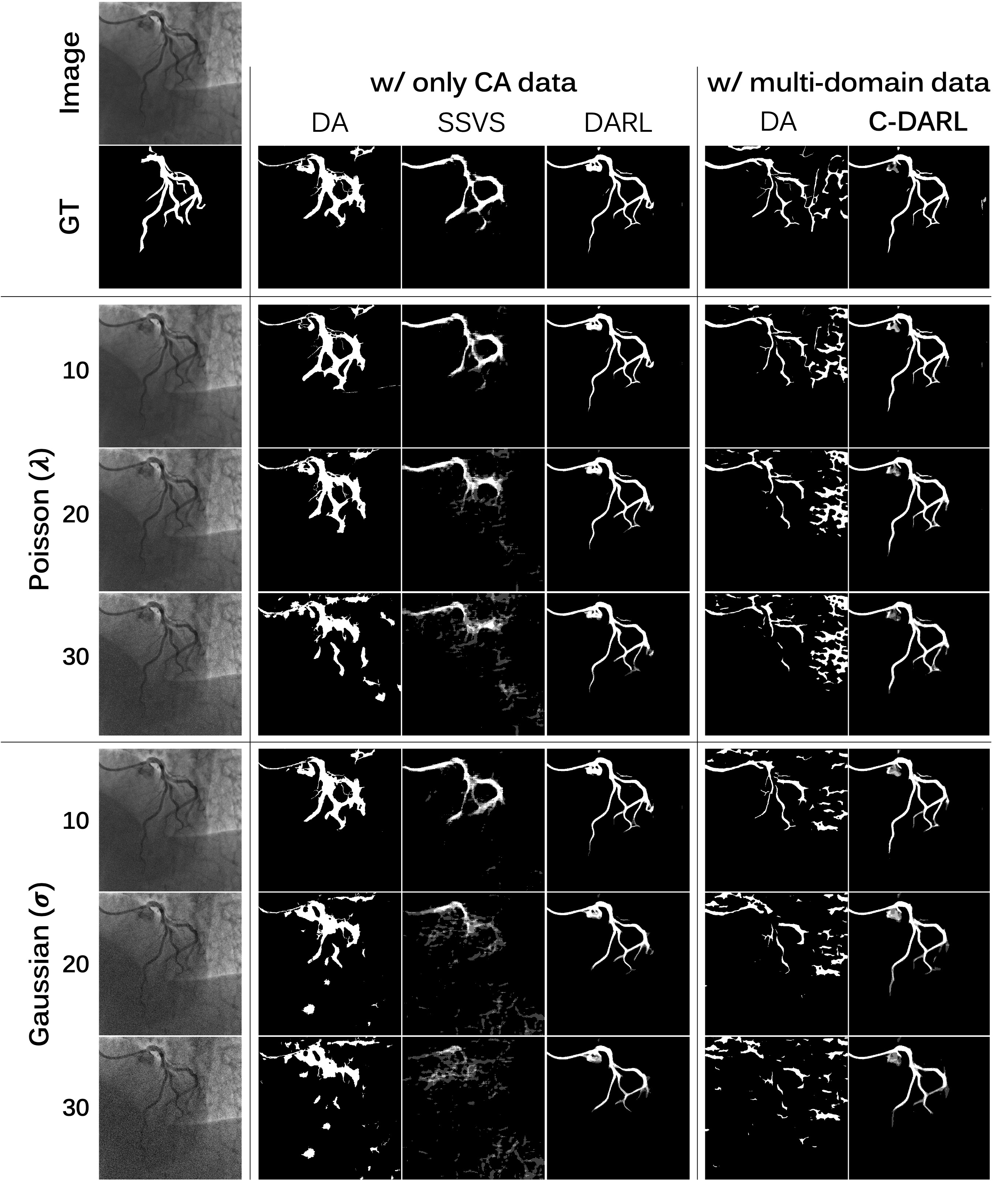}
\caption{Visual vessel segmentation results of the noisy corruption study.}
\label{fig:noise}
\end{figure}

%%%%%%%%%%%%%%%%%%%%%%%%%%%%%%%%%%%%%%%%%%%%%%%%%%%%%%%%%
\subsubsection{Noise corruption study}
\label{sec: noise}
In clinical practice, X-ray angiography images can be degraded by noise due to various factors such as low radiation dose, body mass, organ or patient motion, breathing, or X-ray energy parameters in the data acquisition procedure. Accordingly, we further evaluate the segmentation performance of our C-DARL under noise degradation scenarios. To simulate the noisy images, we add Poisson and Gaussian noises with different levels of $\lambda$ and $\sigma$, respectively, to the clean XCAD test data. Then, we extract vessel structures within the noise-degraded images. 

Fig.~\ref{fig:compare_noise} and Fig.~\ref{fig:noise} show the quantitative and qualitative results of the XCAD test data between our model and the baseline methods. The detailed values of evaluation metrics are reported in Table~\ref{tab:supple_noise} of Supplementary Material. We can observe that our model is robust to noise compared to the existing models even though the performance decreases according to the noise level being stronger. Also, the DARL and C-DARL are the only methods that endure harsh noise corruption with reasonable performance. 
These results suggest that our proposed diffusion adversarial learning framework is superior in unseen noise distributions since the model is trained with highly perturbed input images through the diffusion module.

%%%%%%%%%%%%%%%%%%%%%%%%%%%%%%%%%%%%%%%%%%%%%%%%%%%%%%%%%%%%%%%%%%%%%%%%%%%%
\subsubsection{Ablation study}

Our proposed method is trained by minimizing the objective function (\ref{eq:loss}) which is composed of diffusion loss, contrastive loss, adversarial loss, and cycle loss. To verify the effectiveness of each loss, we conduct an ablation study on the loss function and report the results in Table~\ref{tab:ablation}. 

When our C-DARL model is trained without the diffusion loss (w/o $\mathcal{L}_{diff}$), the segmentation performance slightly decreases than the proposed method. This suggests that the DDPM network architecture is effective to extract features for given noisy input images but learning data distribution through the diffusion loss, which enhances the performance of semantic image synthesis required to capture vessel structures, gives more improvement in vessel segmentation performance. 

\begin{table}[!t]
    \caption{Blood vessel segmentation results for the XCAD test dataset in the ablation study.} 
    \centering
    \resizebox{0.85\linewidth}{!}{
    \begin{tabular}{lwc{2cm}wc{2cm}wc{2cm}}

      \toprule
      \multirow{2}{*}{\textbf{Ablation method}} & \multicolumn{3}{c}{\textbf{Metric}} \\
      \cmidrule(l){2-4}
      & IoU & Dice & Precision \\ 
      \midrule
        w/o $\mathcal{L}_{diff}$ & $0.483 _{\pm 0.094}$ & $0.646 _{\pm 0.088}$ & $0.741 _{\pm 0.114}$ \\
        w/o $\mathcal{L}_{cont}$ & $0.361 _{\pm 0.067}$ & $0.527 _{\pm 0.073}$ & $0.593 _{\pm 0.101}$ \\
        w/o $\mathcal{L}_{adv}$ & $0.004 _{\pm 0.005}$ & $0.008 _{\pm 0.010}$ & $0.006 _{\pm 0.008}$ \\
        w/o $\mathcal{L}_{cyc}$ & $0.091 _{\pm 0.038}$ & $0.164 _{\pm 0.063}$ & $0.107 _{\pm 0.045}$ \\
        Replace $\mathcal{L}_{cont}\rightarrow \mathcal{L}_{adv}$ & $0.456 _{\pm 0.098}$ & $0.620 _{\pm 0.095}$ & $0.669 _{\pm 0.134}$ \\
        w/o \textit{cat} before $G_s$ & $0.489 _{\pm 0.090}$ & $0.652 _{\pm 0.084}$ & $0.717 _{\pm 0.115}$ \\
        \midrule
        Proposed & $\bf0.498_{\pm 0.086}$ & $\bf0.661_{\pm 0.079}$ & $\bf0.750_{\pm 0.108}$ \\
      \bottomrule
     
    \end{tabular}
    }
    \label{tab:ablation}
    \vspace{-0.2cm}
\end{table}

Also, the ablation method that excludes the contrastive loss (w/o $\mathcal{L}_{cont}$) indicates that the proposed mask-based contrastive loss makes the model effectively learn vessel representations without ground-truth labels. In particular, although self-supervised learning can be possible via adversarial learning on the real and fake vessel masks, the experiment in which the contrastive loss is replaced with the adversarial loss (Replace $\mathcal{L}_{cont}\rightarrow \mathcal{L}_{adv}$), such as the loss function of the DARL, have 4\% lower IoU and Dice scores and 8\% lower Precision than the proposed method. This shows that our contrastive loss considering that the input fractal masks have different features from real blood vessels outperforms the adversarial loss regarding the synthetic fractal masks as real masks.

For the C-DARL model trained without the adversarial loss (w/o $\mathcal{L}_{adv}$), the vessel structures are hardly extracted due to no guidance to learn vessel representations. Since the fake vessel images synthesized using the fractal mask provide vessel-lie structure information to the model in the cyclic contrastive path, the adversarial loss is required to generate realistic vessel images. In addition, the model trained without the cycle loss shows much inferior vessel segmentation performance compared to the proposed method, implying that the cycle loss in the cyclic contrastive path enables the model to capture vessel structures of input images.

On the other hand, we also study the necessity of input image concatenation before the generation module. We implement the ablation method that excludes the concatenation (w/o cat before $G_s$), by setting the contrastive path of the generation module to take only noisy vessel images while the adversarial path to take only latent images estimated from the diffusion module. As a result, the segmentation performance is similar to the proposed C-DARL model, but this ablation method shows relatively higher false positives. This indicates that although the two paths of the generation module may require different image information, the latent features from the diffusion module and the noisy vessel images synergistically improve the vessel segmentation performance and their concatenation further simplify the model flows.

%%%%%%%%%%%%%%%%%%%%%%%%%%%%%%%%%%%%%%%%%%%%%%%%%%%%%%%%%%%%%%%%%%%%%%%%%%%%
\subsubsection{Hyperparameter setting}

\begin{figure}[!t]
\centering
\includegraphics[width=\linewidth]{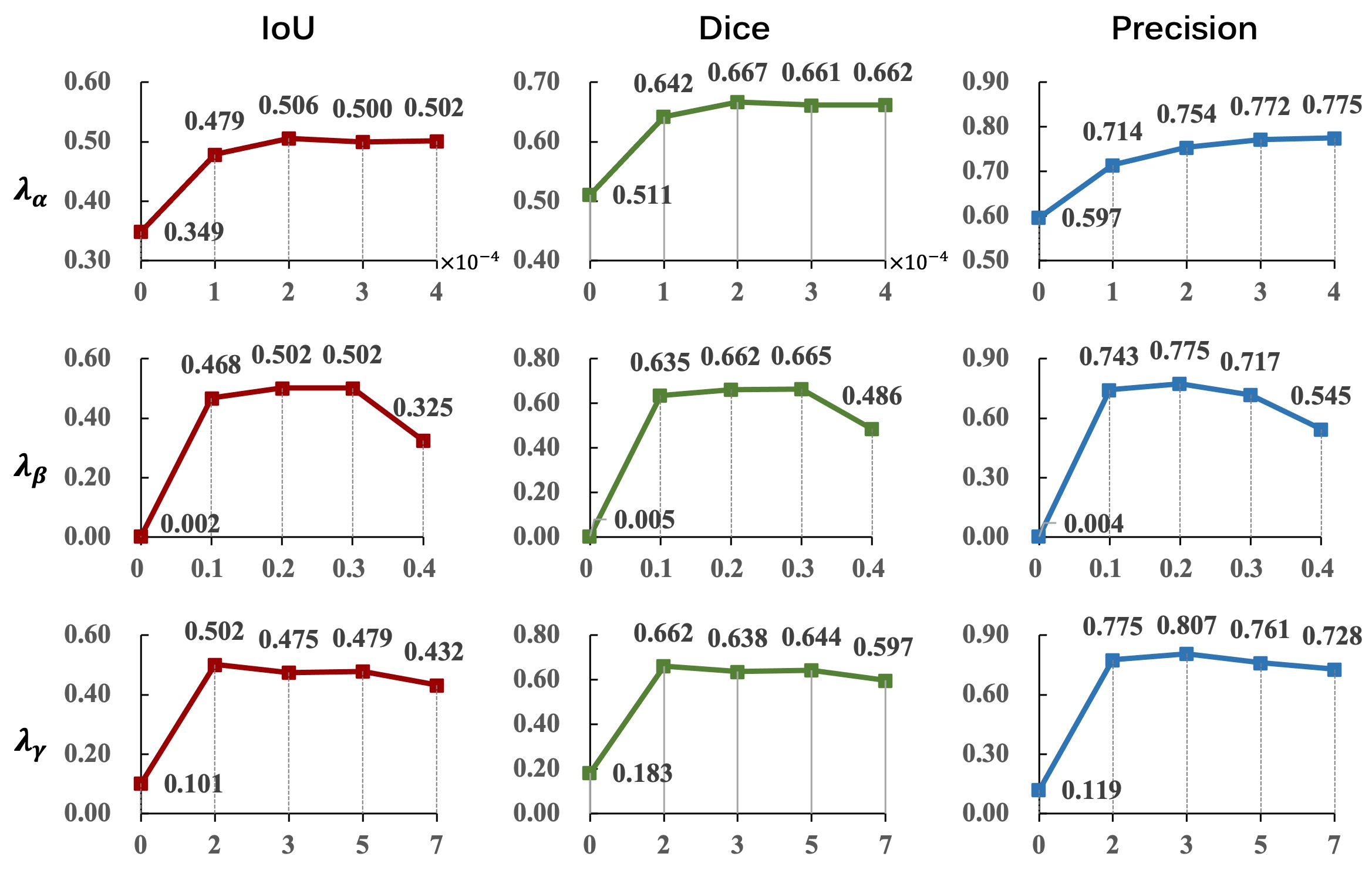}
\caption{Study of hyperparameters in the loss function of C-DARL model. We compare the evaluation results on the validation dataset according to the value of hyperparameters for $\lambda_\alpha$ (top row), $\lambda_\beta$ (middle row), and $\lambda_\gamma$ (bottom row).}
\label{fig:disc_hyper}
\end{figure}

To investigate the optimal setting of hyper-parameters in our loss function (~\ref{eq:loss}), we implement our model by adjusting the values for each $\lambda_\alpha$, $\lambda_\beta$, and $\lambda_\gamma$. When one of the parameters is adjusted, we set the other parameters to fixed values. Fig.~\ref{fig:disc_hyper} shows quantitative evaluation results on the validation dataset according to the hyperparameters.

When the parameter $\lambda_\alpha$ weighting the contrastive loss increases, the IoU and Dice scores increase and then converge, and the Precision value continues to increase slightly, leading to the achievement of the highest performance when $\lambda_\alpha$ is $4\times 10^{-4}$. In the study of $\lambda_\beta$ that controls the adversarial loss, we can observe that our model hardly captures vessel information if the model does not learn semantic image synthesis with $\lambda_\beta=0$, but $\lambda_\beta$ is between 0.1 and 0.3, the model learns vessel representation with high performance. In particular, when $\lambda_\beta$ is set to 0.2, the model shows the best results in terms of IoU and Precision. Lastly, in the study of $\lambda_\gamma$ for the cycle loss, the model shows the best performance at $\lambda_\gamma=2.0$, and the performance gradually decreases as the parameter increases. By considering these results, we report the performance of our model that is trained by setting the hyperparameters as $\lambda_\alpha=4\times 10^{-4}$, $\lambda_\beta=0.2$, and $\lambda_\gamma=2.0$.

%%%%%%%%%%%%%%%%%%%%%%%%%%%%%%%%%%%%%%%%%%%%%%%%%%%%%%%%%%%%%%%%%%%%%%%%%%%%
%%%%%%%%%%%%%%%%%%%%%%%%%%%%%%%%%%%%%%%%%%%%%%%%%%%%%%%%%%%%%%%%%%%%%%%%%%%%
\section{Discussion}

%%%%%%%%%%%%%%%%%%%%%%%%%%%%%%%%%%%%%%%%%%%%%%%%%%%%%%%%%%%%%%%%%%%%%%%%%%%%
\subsection{Training using multi-domain data}

\begin{table}[!t]
    \caption{Quantitative evaluation results of vessel segmentation performance for the comparison study with the self-supervised methods.} 
    \centering
    \resizebox{0.9\linewidth}{!}{
    \begin{tabular}{wl{1.5cm}wl{1.8cm}wc{2.cm}wc{2.cm}wc{2.2cm}}
     \toprule
        \multirow{2}{*}{\bf{Datset}} & \multirow{2}{*}{\bf{Metric}} & \multicolumn{3}{c}{\bf{Training dataset}} \\
        \cmidrule(l){3-5} 
        & & CA & CA+RI & CA+AA+RI  \\ 
        \midrule
        
        \multicolumn{5}{l}{Coronary angiography (CA)} \\ \midrule
        \multirow{3}{*}{XCAD} & IoU & $0.466_{\pm 0.073}$ & $0.491_{\pm 0.088}$ & $\bf0.498_{\pm 0.086}$ \\ 
        & Dice & $0.633_{\pm 0.070}$ & $0.654_{\pm 0.081}$ & $\bf0.661_{\pm 0.079}$ \\ 
        & Precision & $0.712_{\pm 0.114}$ & $0.745_{\pm 0.111}$ & $\bf0.750_{\pm 0.108}$ \\ 
         \cmidrule(l){1-1} \cmidrule(l){2-2} \cmidrule(l){3-5} 

        \multirow{3}{*}{134XCA} & IoU & $0.508_{\pm 0.103}$ & $0.521_{\pm 0.106}$ & $\bf0.545_{\pm 0.094}$ \\ 
        & Dice & $0.668_{\pm 0.097}$ & $0.678_{\pm 0.101}$ & {$\bf0.700_{\pm 0.087}$} \\ 
        & Precision & $0.665_{\pm 0.119}$ & $0.649_{\pm 0.120}$ & {$\bf0.673_{\pm 0.128}$} \\ 
        \cmidrule(l){1-1} \cmidrule(l){2-2} \cmidrule(l){3-5} 

        \multirow{3}{*}{30XCA} & IoU & $0.429_{\pm 0.054}$ & $\bf0.464_{\pm 0.057}$ & {$0.453_{\pm 0.061}$} \\ 
        & Dice & $0.598_{\pm 0.053}$ & $\bf0.632_{\pm 0.053}$ & {$0.621_{\pm 0.059}$} \\ 
        & Precision & $0.806_{\pm 0.105}$ & $0.817_{\pm 0.093}$ & {$\bf0.826_{\pm 0.083}$} \\ \midrule

        \multicolumn{5}{l}{Retinal imaging (RI)} \\ \midrule
        \multirow{3}{*}{DRIVE} & IoU & $0.302_{\pm 0.049}$ & $0.314_{\pm 0.065}$ & $\bf0.345_{\pm 0.061}$ \\ 
        & Dice & $0.462_{\pm 0.058}$ & $0.475_{\pm 0.073}$ & $\bf0.510_{\pm 0.067}$ \\ 
        & Precision & $0.838_{\pm 0.102}$ & $0.888_{\pm 0.103}$ & $\bf0.904_{\pm 0.090}$ \\ 
        \cmidrule(l){1-1} \cmidrule(l){2-2} \cmidrule(l){3-5} 
        
        \multirow{3}{*}{STARE} & IoU & $0.333_{\pm 0.131}$ & $0.342_{\pm 0.117}$ & $\bf0.367_{\pm 0.133}$ \\ 
        & Dice & $0.484_{\pm 0.157}$ & $0.498_{\pm 0.137}$ & $\bf0.522_{\pm 0.154}$ \\ 
        & Precision & $0.685_{\pm 0.190}$ & $0.691_{\pm 0.179}$ & $\bf0.736_{\pm 0.179}$ \\  
        \bottomrule
      \multicolumn{5}{l}{CA: Coronary angiography, AA: Abdoman angiogrpahy, RI: Retinal imaging} 

    \end{tabular}
    }\label{tab:disc_dataset}
    \vspace{-0.2cm}
\end{table}

One of the strengths of our framework is that only vessel images are taken as input without background images before the contrast agent injection so that the model can utilize various vessel medical data, which improves the generalization performance. In order to demonstrate the effect of using multi-domain data in the training phase, we train our C-DARL model with respect to the number of vessel image domains. 

Table~\ref{tab:disc_dataset} reports the vessel segmentation performance according to the input image datasets. When compared to the model trained only using coronary angiography images from the XCAD dataset, the model additionally leveraging vessel data in different domains improves the segmentation performance with a gain from 3\% to 5\% for both internal and external test datasets. Also, the model trained using various datasets in three domains, including coronary images from the XCAD dataset, abdomen angiography images from the AAIR dataset, and retinal images from the UWF and FP datasets, achieves the highest performance in most of the test data. In other words, the segmentation performance increases even though the training data have different domain information, which suggests that our model outperforms to incorporate various domains and achieves better generalization. Future work will study performance for different clinical tasks in scenarios where the underlying vessel pathway is desirable (navigation tasks), as well as settings where diagnostic surrogates for vessel or endothelial pathology are valuable.

%%%%%%%%%%%%%%%%%%%%%%%%%%%%%%%%%%%%%%%%%%%%%%%%%%%%%%%%%%%%%%%%%%%%%%%%%%%%
\subsection{Noise level for inference}
\label{sec:dis_nlevel}

Our framework has another strong point in providing robust performance under noise corruption scenarios thanks to the proposed diffusion adversarial learning strategy, which leverages both the perturbed real vessel image through the forward diffusion process and the synthesized fake vessel image for learning blood vessel segmentation. We analyze the segmentation performance according to the noise level at a time step $t \in[0, 200]$ that is uniformly sampled with an interval of 20. 

In Fig.~\ref{fig:disc_infer}, our model achieves the reliable performance of vessel segmentation until $t$ reaches 200, which relates to Section~\ref{sec: noise} which shows the additional noise corruption study with different types of noise on external datasets. Although the model reveals the best segmentation performance at $t=0$, this inference study with different levels of noise demonstrates the great potential of our model in generalizing to out-of-distribution datasets.

\begin{figure}[!t]
\centering
\includegraphics[width=0.75\linewidth]{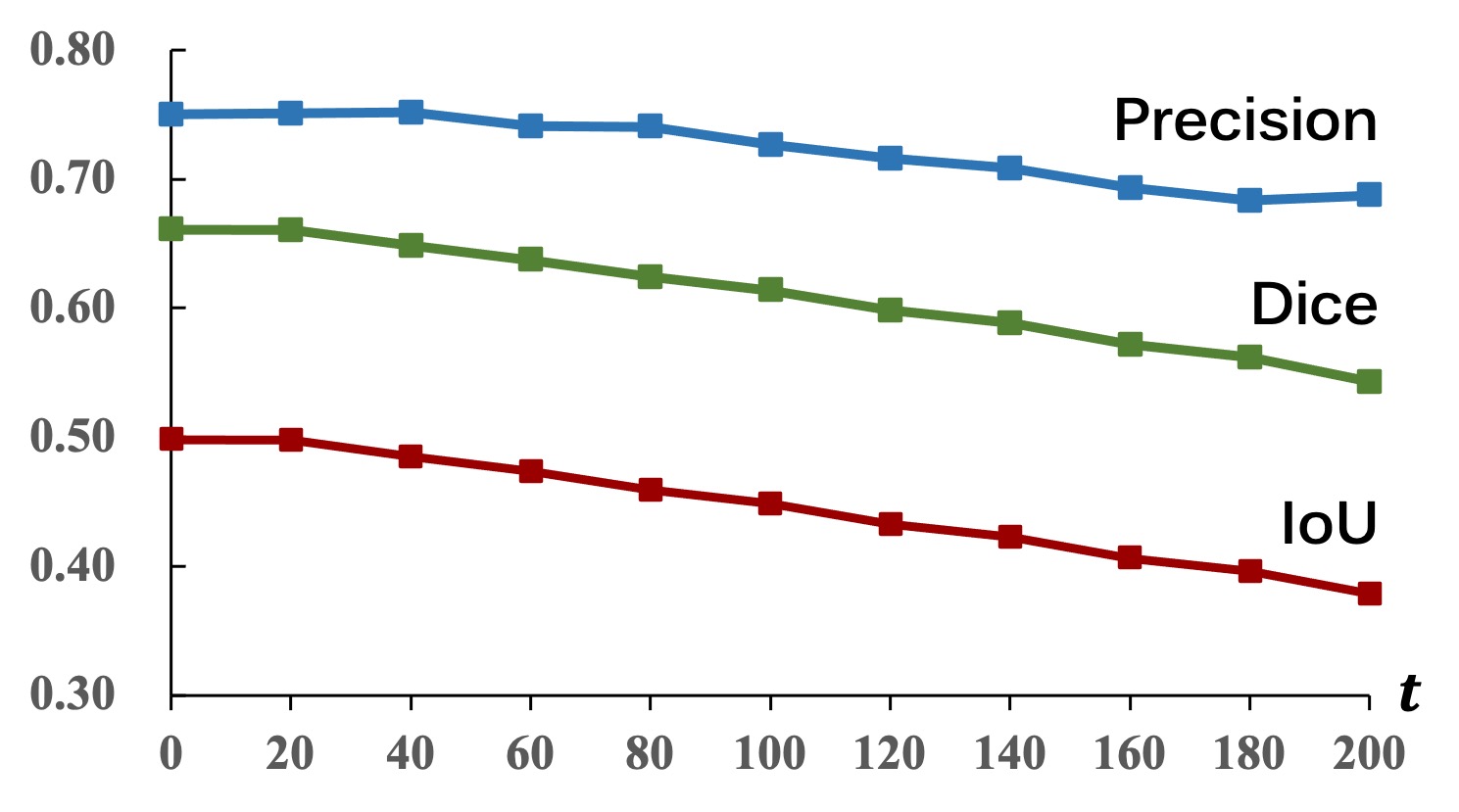}
\caption{The quantitative results of our model on the validation set according to the noise level in the inference phase. The graph shows the value of each evaluation metric (y-axis) with respect to the noise level $t$ (x-axis).}
\label{fig:disc_infer}
\end{figure}

\begin{figure}[!t]
\centering
\includegraphics[width=\linewidth]{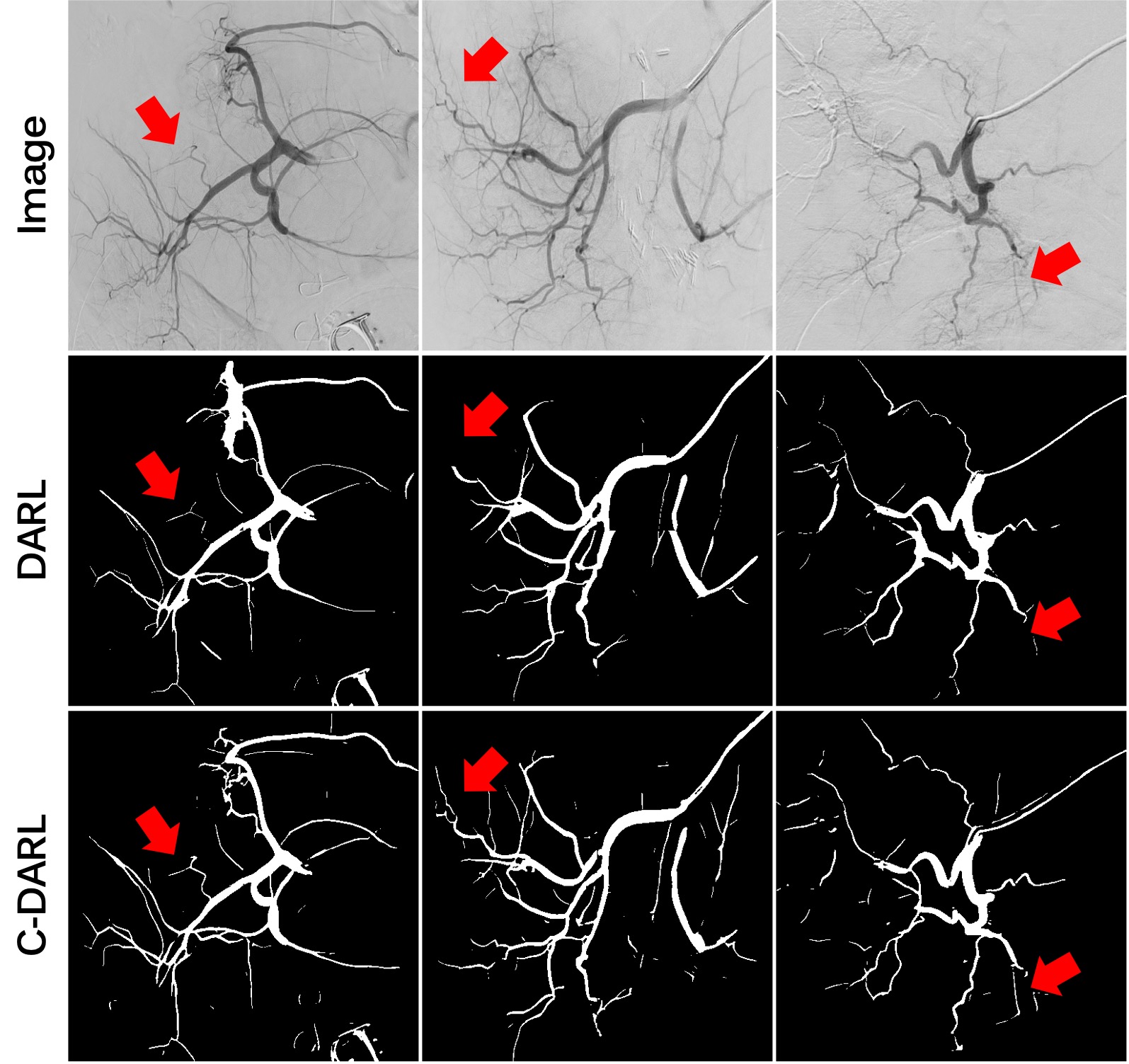}
\caption{Visual vessel segmentation results of abdomen angiography dataset. Red arrows point to remarkable regions.}
\label{fig:disc_aa}
\end{figure}

%%%%%%%%%%%%%%%%%%%%%%%%%%%%%%%%%%%%%%%%%%%%%%%%%%%%%%%%%%%%%%%%%%%%%%%%%%%%
\subsection{Abdomen angiography segmentation}
We also verify the segmentation performance of our model for abdominal angiography to further study the generalization of our model to various domain datasets. As there is no benchmark dataset containing ground-truth vessel labels for the abdominal vessel images, we instead compare the qualitative results of the vessel segmentation using the AAIR data. For a baseline model, we adopt the DARL model which achieves the second-best performance in the comparison study of Section~\ref{sec:result_comp} (Table~\ref{tab:compare_self}). 

Fig.~\ref{fig:disc_aa} shows the results of the blood vessel segmentation for several abdomen angiography images. Compared with the DARL model, the results show that the proposed C-DARL model outperforms the segmentation of vascular regions, including tiny and low-contrast branches. Also, our model shows superior segmentation performance of even blood vessels that are difficult to distinguish from the background structures. This suggests that our model can be applied to a variety of vascular images by improving the capability of learning vessel representations through contrastive learning under the condition of no ground-truth vessel masks.

%%%%%%%%%%%%%%%%%%%%%%%%%%%%%%%%%%%%%%%%%%%%%%%%%%%%%%%%%%%%%%%%%%%%%%%%%%%%
%%%%%%%%%%%%%%%%%%%%%%%%%%%%%%%%%%%%%%%%%%%%%%%%%%%%%%%%%%%%%%%%%%%%%%%%%%%%
\section{Conclusion}
We propose a novel label-free C-DARL model that achieves reliable blood vessel segmentation performance on multi-domain vessel images. Our model addresses the limitations of previously introduced DARL by effectively simplifying two different vessel image generation and vessel segmentation tasks, enabling self-supervised learning of blood vessel features using vessel images in a variety of domains. We further introduce the mask-based contrastive learning to the diffusion adversarial learning framework, by leveraging both the synthetic fractal and the estimated segmentation masks to intensely extract vessel representation under the label-free condition. Extensive experiments conducted on various vessel segmentation benchmarks demonstrate that our proposed C-DARL outperforms existing label-free vessel segmentation methods, offering noise robustness and promising generalization performance across diverse vessel datasets. This performance with a variety of clinical settings for vessel segmentation suggests that our model can provide clinicians with presumptive guidance during the diagnostic process for a variety of organs and diseases, as well as aid in detection, treatment planning, navigation, or reference data augmentation for a variety of clinical interventional and diagnostic tasks %various diseases or aid in planning surgical therapy 
without requiring any labeling process.

\section*{Acknowledgments}
This research was supported in part by the Intramural Research Program of the National Institutes of Health, Clinical Center, United States, and in part by the National Research Foundation of Korea under Grant
NRF-2020R1A2B5B03001980.}

\bibliographystyle{model2-names.bst}\biboptions{authoryear}
\bibliography{MEDIA_DCRL_ref}

\begin{thebibliography}{41}
\expandafter\ifx\csname natexlab\endcsname\relax\def\natexlab#1{#1}\fi
\providecommand{\url}[1]{\texttt{#1}}
\providecommand{\href}[2]{#2}
\providecommand{\path}[1]{#1}
\providecommand{\DOIprefix}{doi:}
\providecommand{\ArXivprefix}{arXiv:}
\providecommand{\URLprefix}{URL: }
\providecommand{\Pubmedprefix}{pmid:}
\providecommand{\doi}[1]{\href{http://dx.doi.org/#1}{\path{#1}}}
\providecommand{\Pubmed}[1]{\href{pmid:#1}{\path{#1}}}
\providecommand{\bibinfo}[2]{#2}
\ifx\xfnm\relax \def\xfnm[#1]{\unskip,\space#1}\fi
%Type = Inproceedings
\bibitem[{Baranchuk et~al.(2021)Baranchuk, Voynov, Rubachev, Khrulkov and
  Babenko}]{baranchuk2021label}
\bibinfo{author}{Baranchuk, D.}, \bibinfo{author}{Voynov, A.},
  \bibinfo{author}{Rubachev, I.}, \bibinfo{author}{Khrulkov, V.},
  \bibinfo{author}{Babenko, A.}, \bibinfo{year}{2021}.
\newblock \bibinfo{title}{Label-efficient semantic segmentation with diffusion
  models}, in: \bibinfo{booktitle}{International Conference on Learning
  Representations}.
%Type = Article
\bibitem[{Brempong et~al.(2022)Brempong, Kornblith, Chen, Parmar, Minderer and
  Norouzi}]{asiedu2022decoder}
\bibinfo{author}{Brempong, E.A.}, \bibinfo{author}{Kornblith, S.},
  \bibinfo{author}{Chen, T.}, \bibinfo{author}{Parmar, N.},
  \bibinfo{author}{Minderer, M.}, \bibinfo{author}{Norouzi, M.},
  \bibinfo{year}{2022}.
\newblock \bibinfo{title}{Decoder denoising pretraining for semantic
  segmentation}.
\newblock \bibinfo{journal}{Transactions on Machine Learning Research}
  \URLprefix \url{https://openreview.net/forum?id=D3WI0QG7dC}.
%Type = Article
\bibitem[{Cervantes-Sanchez et~al.(2019)Cervantes-Sanchez, Cruz-Aceves,
  Hernandez-Aguirre, Hernandez-Gonzalez and Solorio-Meza}]{app9245507}
\bibinfo{author}{Cervantes-Sanchez, F.}, \bibinfo{author}{Cruz-Aceves, I.},
  \bibinfo{author}{Hernandez-Aguirre, A.}, \bibinfo{author}{Hernandez-Gonzalez,
  M.A.}, \bibinfo{author}{Solorio-Meza, S.E.}, \bibinfo{year}{2019}.
\newblock \bibinfo{title}{Automatic segmentation of coronary arteries in x-ray
  angiograms using multiscale analysis and artificial neural networks}.
\newblock \bibinfo{journal}{Applied Sciences} \bibinfo{volume}{9}.
\newblock \URLprefix \url{https://www.mdpi.com/2076-3417/9/24/5507},
  \DOIprefix\doi{10.3390/app9245507}.
%Type = Inproceedings
\bibitem[{Chen et~al.(2020)Chen, Kornblith, Norouzi and
  Hinton}]{chen2020simple}
\bibinfo{author}{Chen, T.}, \bibinfo{author}{Kornblith, S.},
  \bibinfo{author}{Norouzi, M.}, \bibinfo{author}{Hinton, G.},
  \bibinfo{year}{2020}.
\newblock \bibinfo{title}{A simple framework for contrastive learning of visual
  representations}, in: \bibinfo{booktitle}{International conference on machine
  learning}, \bibinfo{organization}{PMLR}. pp. \bibinfo{pages}{1597--1607}.
%Type = Article
\bibitem[{Dehkordi et~al.(2011)Dehkordi, Sadri and
  Doosthoseini}]{dehkordi2011review}
\bibinfo{author}{Dehkordi, M.T.}, \bibinfo{author}{Sadri, S.},
  \bibinfo{author}{Doosthoseini, A.}, \bibinfo{year}{2011}.
\newblock \bibinfo{title}{A review of coronary vessel segmentation algorithms}.
\newblock \bibinfo{journal}{Journal of medical signals and sensors}
  \bibinfo{volume}{1}, \bibinfo{pages}{49}.
%Type = Article
\bibitem[{Delibasis et~al.(2010)Delibasis, Kechriniotis, Tsonos and
  Assimakis}]{delibasis2010automatic}
\bibinfo{author}{Delibasis, K.K.}, \bibinfo{author}{Kechriniotis, A.I.},
  \bibinfo{author}{Tsonos, C.}, \bibinfo{author}{Assimakis, N.},
  \bibinfo{year}{2010}.
\newblock \bibinfo{title}{Automatic model-based tracing algorithm for vessel
  segmentation and diameter estimation}.
\newblock \bibinfo{journal}{Computer methods and programs in biomedicine}
  \bibinfo{volume}{100}, \bibinfo{pages}{108--122}.
%Type = Article
\bibitem[{Fan et~al.(2018)Fan, Yang, Wang, Yang, Ai, Huang, Song, Hao and
  Wang}]{fan2018multichannel}
\bibinfo{author}{Fan, J.}, \bibinfo{author}{Yang, J.}, \bibinfo{author}{Wang,
  Y.}, \bibinfo{author}{Yang, S.}, \bibinfo{author}{Ai, D.},
  \bibinfo{author}{Huang, Y.}, \bibinfo{author}{Song, H.},
  \bibinfo{author}{Hao, A.}, \bibinfo{author}{Wang, Y.}, \bibinfo{year}{2018}.
\newblock \bibinfo{title}{Multichannel fully convolutional network for coronary
  artery segmentation in x-ray angiograms}.
\newblock \bibinfo{journal}{Ieee Access} \bibinfo{volume}{6},
  \bibinfo{pages}{44635--44643}.
%Type = Article
\bibitem[{Fan et~al.(2019)Fan, Mo, Qiu, Li, Zhu, Li, Hu, Rong and
  Chen}]{fan2019accurate}
\bibinfo{author}{Fan, Z.}, \bibinfo{author}{Mo, J.}, \bibinfo{author}{Qiu, B.},
  \bibinfo{author}{Li, W.}, \bibinfo{author}{Zhu, G.}, \bibinfo{author}{Li,
  C.}, \bibinfo{author}{Hu, J.}, \bibinfo{author}{Rong, Y.},
  \bibinfo{author}{Chen, X.}, \bibinfo{year}{2019}.
\newblock \bibinfo{title}{Accurate retinal vessel segmentation via octave
  convolution neural network}.
\newblock \bibinfo{journal}{arXiv preprint arXiv:1906.12193} .
%Type = Article
\bibitem[{Hao et~al.(2020)Hao, Ding, Qiu, Lv, Fei, Zhu and Qin}]{HAO2020172}
\bibinfo{author}{Hao, D.}, \bibinfo{author}{Ding, S.}, \bibinfo{author}{Qiu,
  L.}, \bibinfo{author}{Lv, Y.}, \bibinfo{author}{Fei, B.},
  \bibinfo{author}{Zhu, Y.}, \bibinfo{author}{Qin, B.}, \bibinfo{year}{2020}.
\newblock \bibinfo{title}{Sequential vessel segmentation via deep channel
  attention network}.
\newblock \bibinfo{journal}{Neural Networks} \bibinfo{volume}{128},
  \bibinfo{pages}{172--187}.
\newblock \URLprefix
  \url{https://www.sciencedirect.com/science/article/pii/S0893608020301672},
  \DOIprefix\doi{https://doi.org/10.1016/j.neunet.2020.05.005}.
%Type = Inproceedings
\bibitem[{He et~al.(2020)He, Fan, Wu, Xie and Girshick}]{he2020momentum}
\bibinfo{author}{He, K.}, \bibinfo{author}{Fan, H.}, \bibinfo{author}{Wu, Y.},
  \bibinfo{author}{Xie, S.}, \bibinfo{author}{Girshick, R.},
  \bibinfo{year}{2020}.
\newblock \bibinfo{title}{Momentum contrast for unsupervised visual
  representation learning}, in: \bibinfo{booktitle}{Proceedings of the IEEE/CVF
  conference on computer vision and pattern recognition}, pp.
  \bibinfo{pages}{9729--9738}.
%Type = Article
\bibitem[{Ho et~al.(2020)Ho, Jain and Abbeel}]{ho2020denoising}
\bibinfo{author}{Ho, J.}, \bibinfo{author}{Jain, A.}, \bibinfo{author}{Abbeel,
  P.}, \bibinfo{year}{2020}.
\newblock \bibinfo{title}{Denoising diffusion probabilistic models}.
\newblock \bibinfo{journal}{Advances in Neural Information Processing Systems}
  \bibinfo{volume}{33}, \bibinfo{pages}{6840--6851}.
%Type = Article
\bibitem[{Hoover et~al.(2000)Hoover, Kouznetsova and
  Goldbaum}]{hoover2000locating}
\bibinfo{author}{Hoover, A.}, \bibinfo{author}{Kouznetsova, V.},
  \bibinfo{author}{Goldbaum, M.}, \bibinfo{year}{2000}.
\newblock \bibinfo{title}{Locating blood vessels in retinal images by piecewise
  threshold probing of a matched filter response}.
\newblock \bibinfo{journal}{IEEE Transactions on Medical imaging}
  \bibinfo{volume}{19}, \bibinfo{pages}{203--210}.
%Type = Inproceedings
\bibitem[{Hu et~al.(2021)Hu, Cui and Wang}]{Hu_2021_ICCV}
\bibinfo{author}{Hu, H.}, \bibinfo{author}{Cui, J.}, \bibinfo{author}{Wang,
  L.}, \bibinfo{year}{2021}.
\newblock \bibinfo{title}{Region-aware contrastive learning for semantic
  segmentation}, in: \bibinfo{booktitle}{Proceedings of the IEEE/CVF
  International Conference on Computer Vision (ICCV)}, pp.
  \bibinfo{pages}{16291--16301}.
%Type = Inproceedings
\bibitem[{Huang et~al.(2023)Huang, Chan, Jiang and Liu}]{Huang_2023_CVPR}
\bibinfo{author}{Huang, Z.}, \bibinfo{author}{Chan, K.C.},
  \bibinfo{author}{Jiang, Y.}, \bibinfo{author}{Liu, Z.}, \bibinfo{year}{2023}.
\newblock \bibinfo{title}{Collaborative diffusion for multi-modal face
  generation and editing}, in: \bibinfo{booktitle}{Proceedings of the IEEE/CVF
  Conference on Computer Vision and Pattern Recognition (CVPR)}, pp.
  \bibinfo{pages}{6080--6090}.
%Type = Inproceedings
\bibitem[{Isola et~al.(2017)Isola, Zhu, Zhou and Efros}]{isola2017image}
\bibinfo{author}{Isola, P.}, \bibinfo{author}{Zhu, J.Y.},
  \bibinfo{author}{Zhou, T.}, \bibinfo{author}{Efros, A.A.},
  \bibinfo{year}{2017}.
\newblock \bibinfo{title}{Image-to-image translation with conditional
  adversarial networks}, in: \bibinfo{booktitle}{Proceedings of the IEEE
  conference on computer vision and pattern recognition}, pp.
  \bibinfo{pages}{1125--1134}.
%Type = Article
\bibitem[{Jiang et~al.(2019)Jiang, Zhang, Tan and Chen}]{jiang2019automatic}
\bibinfo{author}{Jiang, Y.}, \bibinfo{author}{Zhang, H.}, \bibinfo{author}{Tan,
  N.}, \bibinfo{author}{Chen, L.}, \bibinfo{year}{2019}.
\newblock \bibinfo{title}{Automatic retinal blood vessel segmentation based on
  fully convolutional neural networks}.
\newblock \bibinfo{journal}{Symmetry} \bibinfo{volume}{11},
  \bibinfo{pages}{1112}.
%Type = Article
\bibitem[{Kim et~al.(2022)Kim, Oh and Ye}]{kim2022diffusion}
\bibinfo{author}{Kim, B.}, \bibinfo{author}{Oh, Y.}, \bibinfo{author}{Ye,
  J.C.}, \bibinfo{year}{2022}.
\newblock \bibinfo{title}{Diffusion adversarial representation learning for
  self-supervised vessel segmentation}.
\newblock \bibinfo{journal}{arXiv preprint arXiv:2209.14566} .
%Type = Article
\bibitem[{Kingma and Ba(2014)}]{kingma2014adam}
\bibinfo{author}{Kingma, D.P.}, \bibinfo{author}{Ba, J.}, \bibinfo{year}{2014}.
\newblock \bibinfo{title}{Adam: A method for stochastic optimization}.
\newblock \bibinfo{journal}{arXiv preprint arXiv:1412.6980} .
%Type = Article
\bibitem[{Lesage et~al.(2009)Lesage, Angelini, Bloch and
  Funka-Lea}]{lesage2009review}
\bibinfo{author}{Lesage, D.}, \bibinfo{author}{Angelini, E.D.},
  \bibinfo{author}{Bloch, I.}, \bibinfo{author}{Funka-Lea, G.},
  \bibinfo{year}{2009}.
\newblock \bibinfo{title}{A review of 3d vessel lumen segmentation techniques:
  Models, features and extraction schemes}.
\newblock \bibinfo{journal}{Medical image analysis} \bibinfo{volume}{13},
  \bibinfo{pages}{819--845}.
%Type = Inproceedings
\bibitem[{Ma et~al.(2021)Ma, Hua, Deng, Song, Wang, Xue, Cao, Ma and
  Guan}]{ma2021self}
\bibinfo{author}{Ma, Y.}, \bibinfo{author}{Hua, Y.}, \bibinfo{author}{Deng,
  H.}, \bibinfo{author}{Song, T.}, \bibinfo{author}{Wang, H.},
  \bibinfo{author}{Xue, Z.}, \bibinfo{author}{Cao, H.}, \bibinfo{author}{Ma,
  R.}, \bibinfo{author}{Guan, H.}, \bibinfo{year}{2021}.
\newblock \bibinfo{title}{Self-supervised vessel segmentation via adversarial
  learning}, in: \bibinfo{booktitle}{Proceedings of the IEEE/CVF International
  Conference on Computer Vision}, pp. \bibinfo{pages}{7536--7545}.
%Type = Article
\bibitem[{Mahmood et~al.(2019)Mahmood, Borders, Chen, McKay, Salimian, Baras
  and Durr}]{mahmood2019deep}
\bibinfo{author}{Mahmood, F.}, \bibinfo{author}{Borders, D.},
  \bibinfo{author}{Chen, R.J.}, \bibinfo{author}{McKay, G.N.},
  \bibinfo{author}{Salimian, K.J.}, \bibinfo{author}{Baras, A.},
  \bibinfo{author}{Durr, N.J.}, \bibinfo{year}{2019}.
\newblock \bibinfo{title}{Deep adversarial training for multi-organ nuclei
  segmentation in histopathology images}.
\newblock \bibinfo{journal}{IEEE transactions on medical imaging}
  \bibinfo{volume}{39}, \bibinfo{pages}{3257--3267}.
%Type = Inproceedings
\bibitem[{Mao et~al.(2017)Mao, Li, Xie, Lau, Wang and
  Paul~Smolley}]{mao2017least}
\bibinfo{author}{Mao, X.}, \bibinfo{author}{Li, Q.}, \bibinfo{author}{Xie, H.},
  \bibinfo{author}{Lau, R.Y.}, \bibinfo{author}{Wang, Z.},
  \bibinfo{author}{Paul~Smolley, S.}, \bibinfo{year}{2017}.
\newblock \bibinfo{title}{Least squares generative adversarial networks}, in:
  \bibinfo{booktitle}{Proceedings of the IEEE international conference on
  computer vision}, pp. \bibinfo{pages}{2794--2802}.
%Type = Inproceedings
\bibitem[{Melas-Kyriazi et~al.(2022)Melas-Kyriazi, Rupprecht, Laina and
  Vedaldi}]{melas2022deep}
\bibinfo{author}{Melas-Kyriazi, L.}, \bibinfo{author}{Rupprecht, C.},
  \bibinfo{author}{Laina, I.}, \bibinfo{author}{Vedaldi, A.},
  \bibinfo{year}{2022}.
\newblock \bibinfo{title}{Deep spectral methods: A surprisingly strong baseline
  for unsupervised semantic segmentation and localization}, in:
  \bibinfo{booktitle}{Proceedings of the IEEE/CVF Conference on Computer Vision
  and Pattern Recognition}, pp. \bibinfo{pages}{8364--8375}.
%Type = Inproceedings
\bibitem[{Nasr-Esfahani et~al.(2016)Nasr-Esfahani, Samavi, Karimi, Soroushmehr,
  Ward, Jafari, Felfeliyan, Nallamothu and Najarian}]{nasr2016vessel}
\bibinfo{author}{Nasr-Esfahani, E.}, \bibinfo{author}{Samavi, S.},
  \bibinfo{author}{Karimi, N.}, \bibinfo{author}{Soroushmehr, S.R.},
  \bibinfo{author}{Ward, K.}, \bibinfo{author}{Jafari, M.H.},
  \bibinfo{author}{Felfeliyan, B.}, \bibinfo{author}{Nallamothu, B.},
  \bibinfo{author}{Najarian, K.}, \bibinfo{year}{2016}.
\newblock \bibinfo{title}{Vessel extraction in x-ray angiograms using deep
  learning}, in: \bibinfo{booktitle}{2016 38th Annual international conference
  of the IEEE engineering in medicine and biology society (EMBC)},
  \bibinfo{organization}{IEEE}. pp. \bibinfo{pages}{643--646}.
%Type = Inproceedings
\bibitem[{Oh et~al.(2022)Oh, Ko and Park}]{Oh_2022_ACCV}
\bibinfo{author}{Oh, Y.}, \bibinfo{author}{Ko, E.S.}, \bibinfo{author}{Park,
  H.}, \bibinfo{year}{2022}.
\newblock \bibinfo{title}{Semi-supervised breast lesion segmentation using
  local cross triplet loss for ultrafast dynamic contrast-enhanced mri}, in:
  \bibinfo{booktitle}{Proceedings of the Asian Conference on Computer Vision
  (ACCV)}, pp. \bibinfo{pages}{2713--2728}.
%Type = Article
\bibitem[{Oord et~al.(2018)Oord, Li and Vinyals}]{oord2018representation}
\bibinfo{author}{Oord, A.v.d.}, \bibinfo{author}{Li, Y.},
  \bibinfo{author}{Vinyals, O.}, \bibinfo{year}{2018}.
\newblock \bibinfo{title}{Representation learning with contrastive predictive
  coding}.
\newblock \bibinfo{journal}{arXiv preprint arXiv:1807.03748} .
%Type = Misc
\bibitem[{Oquab et~al.(2023)Oquab, Darcet, Moutakanni, Vo, Szafraniec,
  Khalidov, Fernandez, Haziza, Massa, El-Nouby, Assran, Ballas, Galuba, Howes,
  Huang, Li, Misra, Rabbat, Sharma, Synnaeve, Xu, Jegou, Mairal, Labatut,
  Joulin and Bojanowski}]{oquab2023dinov2}
\bibinfo{author}{Oquab, M.}, \bibinfo{author}{Darcet, T.},
  \bibinfo{author}{Moutakanni, T.}, \bibinfo{author}{Vo, H.},
  \bibinfo{author}{Szafraniec, M.}, \bibinfo{author}{Khalidov, V.},
  \bibinfo{author}{Fernandez, P.}, \bibinfo{author}{Haziza, D.},
  \bibinfo{author}{Massa, F.}, \bibinfo{author}{El-Nouby, A.},
  \bibinfo{author}{Assran, M.}, \bibinfo{author}{Ballas, N.},
  \bibinfo{author}{Galuba, W.}, \bibinfo{author}{Howes, R.},
  \bibinfo{author}{Huang, P.Y.}, \bibinfo{author}{Li, S.W.},
  \bibinfo{author}{Misra, I.}, \bibinfo{author}{Rabbat, M.},
  \bibinfo{author}{Sharma, V.}, \bibinfo{author}{Synnaeve, G.},
  \bibinfo{author}{Xu, H.}, \bibinfo{author}{Jegou, H.},
  \bibinfo{author}{Mairal, J.}, \bibinfo{author}{Labatut, P.},
  \bibinfo{author}{Joulin, A.}, \bibinfo{author}{Bojanowski, P.},
  \bibinfo{year}{2023}.
\newblock \bibinfo{title}{Dinov2: Learning robust visual features without
  supervision}.
\newblock \href{http://arxiv.org/abs/2304.07193}{\tt arXiv:2304.07193}.
%Type = Inproceedings
\bibitem[{Park et~al.(2020)Park, Efros, Zhang and Zhu}]{park2020contrastive}
\bibinfo{author}{Park, T.}, \bibinfo{author}{Efros, A.A.},
  \bibinfo{author}{Zhang, R.}, \bibinfo{author}{Zhu, J.Y.},
  \bibinfo{year}{2020}.
\newblock \bibinfo{title}{Contrastive learning for unpaired image-to-image
  translation}, in: \bibinfo{booktitle}{Computer Vision--ECCV 2020: 16th
  European Conference, Glasgow, UK, August 23--28, 2020, Proceedings, Part IX
  16}, \bibinfo{organization}{Springer}. pp. \bibinfo{pages}{319--345}.
%Type = Inproceedings
\bibitem[{Park et~al.(2019)Park, Liu, Wang and Zhu}]{park2019semantic}
\bibinfo{author}{Park, T.}, \bibinfo{author}{Liu, M.Y.}, \bibinfo{author}{Wang,
  T.C.}, \bibinfo{author}{Zhu, J.Y.}, \bibinfo{year}{2019}.
\newblock \bibinfo{title}{Semantic image synthesis with spatially-adaptive
  normalization}, in: \bibinfo{booktitle}{Proceedings of the IEEE/CVF
  conference on computer vision and pattern recognition}, pp.
  \bibinfo{pages}{2337--2346}.
%Type = Article
\bibitem[{Paszke et~al.(2019)Paszke, Gross, Massa, Lerer, Bradbury, Chanan,
  Killeen, Lin, Gimelshein, Antiga et~al.}]{paszke2019pytorch}
\bibinfo{author}{Paszke, A.}, \bibinfo{author}{Gross, S.},
  \bibinfo{author}{Massa, F.}, \bibinfo{author}{Lerer, A.},
  \bibinfo{author}{Bradbury, J.}, \bibinfo{author}{Chanan, G.},
  \bibinfo{author}{Killeen, T.}, \bibinfo{author}{Lin, Z.},
  \bibinfo{author}{Gimelshein, N.}, \bibinfo{author}{Antiga, L.}, et~al.,
  \bibinfo{year}{2019}.
\newblock \bibinfo{title}{Pytorch: An imperative style, high-performance deep
  learning library}.
\newblock \bibinfo{journal}{Advances in neural information processing systems}
  \bibinfo{volume}{32}.
%Type = Inproceedings
\bibitem[{Rahman et~al.(2023)Rahman, Valanarasu, Hacihaliloglu and
  Patel}]{Rahman_2023_CVPR}
\bibinfo{author}{Rahman, A.}, \bibinfo{author}{Valanarasu, J.M.J.},
  \bibinfo{author}{Hacihaliloglu, I.}, \bibinfo{author}{Patel, V.M.},
  \bibinfo{year}{2023}.
\newblock \bibinfo{title}{Ambiguous medical image segmentation using diffusion
  models}, in: \bibinfo{booktitle}{Proceedings of the IEEE/CVF Conference on
  Computer Vision and Pattern Recognition (CVPR)}, pp.
  \bibinfo{pages}{11536--11546}.
%Type = Article
\bibitem[{Staal et~al.(2004)Staal, Abr{\`a}moff, Niemeijer, Viergever and
  Van~Ginneken}]{staal2004ridge}
\bibinfo{author}{Staal, J.}, \bibinfo{author}{Abr{\`a}moff, M.D.},
  \bibinfo{author}{Niemeijer, M.}, \bibinfo{author}{Viergever, M.A.},
  \bibinfo{author}{Van~Ginneken, B.}, \bibinfo{year}{2004}.
\newblock \bibinfo{title}{Ridge-based vessel segmentation in color images of
  the retina}.
\newblock \bibinfo{journal}{IEEE transactions on medical imaging}
  \bibinfo{volume}{23}, \bibinfo{pages}{501--509}.
%Type = Article
\bibitem[{Taghizadeh~Dehkordi et~al.(2014)Taghizadeh~Dehkordi, Doost~Hoseini,
  Sadri and Soltanianzadeh}]{taghizadeh2014local}
\bibinfo{author}{Taghizadeh~Dehkordi, M.}, \bibinfo{author}{Doost~Hoseini,
  A.M.}, \bibinfo{author}{Sadri, S.}, \bibinfo{author}{Soltanianzadeh, H.},
  \bibinfo{year}{2014}.
\newblock \bibinfo{title}{Local feature fitting active contour for segmenting
  vessels in angiograms}.
\newblock \bibinfo{journal}{IET Computer Vision} \bibinfo{volume}{8},
  \bibinfo{pages}{161--170}.
%Type = Article
\bibitem[{Ulyanov et~al.(2016)Ulyanov, Vedaldi and
  Lempitsky}]{ulyanov2016instance}
\bibinfo{author}{Ulyanov, D.}, \bibinfo{author}{Vedaldi, A.},
  \bibinfo{author}{Lempitsky, V.}, \bibinfo{year}{2016}.
\newblock \bibinfo{title}{Instance normalization: The missing ingredient for
  fast stylization}.
\newblock \bibinfo{journal}{arXiv preprint arXiv:1607.08022} .
%Type = Misc
\bibitem[{Wang et~al.(2022)Wang, Bao, Zhou, Chen, Chen, Yuan and
  Li}]{wang2022semantic}
\bibinfo{author}{Wang, W.}, \bibinfo{author}{Bao, J.}, \bibinfo{author}{Zhou,
  W.}, \bibinfo{author}{Chen, D.}, \bibinfo{author}{Chen, D.},
  \bibinfo{author}{Yuan, L.}, \bibinfo{author}{Li, H.}, \bibinfo{year}{2022}.
\newblock \bibinfo{title}{Semantic image synthesis via diffusion models}.
\newblock \href{http://arxiv.org/abs/2207.00050}{\tt arXiv:2207.00050}.
%Type = Inproceedings
\bibitem[{Wu et~al.(2019)Wu, Zou and Yang}]{wu2019u}
\bibinfo{author}{Wu, C.}, \bibinfo{author}{Zou, Y.}, \bibinfo{author}{Yang,
  Z.}, \bibinfo{year}{2019}.
\newblock \bibinfo{title}{U-gan: Generative adversarial networks with u-net for
  retinal vessel segmentation}, in: \bibinfo{booktitle}{2019 14th international
  conference on computer science \& education (ICCSE)},
  \bibinfo{organization}{IEEE}. pp. \bibinfo{pages}{642--646}.
%Type = Inproceedings
\bibitem[{Wu et~al.(2018)Wu, Xiong, Yu and Lin}]{wu2018unsupervised}
\bibinfo{author}{Wu, Z.}, \bibinfo{author}{Xiong, Y.}, \bibinfo{author}{Yu,
  S.X.}, \bibinfo{author}{Lin, D.}, \bibinfo{year}{2018}.
\newblock \bibinfo{title}{Unsupervised feature learning via non-parametric
  instance discrimination}, in: \bibinfo{booktitle}{Proceedings of the IEEE
  conference on computer vision and pattern recognition}, pp.
  \bibinfo{pages}{3733--3742}.
%Type = Article
\bibitem[{Yang et~al.(2019)Yang, Kweon, Roh, Lee, Kang, Park, Kim, Yang, Hur,
  Kang et~al.}]{yang2019deep}
\bibinfo{author}{Yang, S.}, \bibinfo{author}{Kweon, J.}, \bibinfo{author}{Roh,
  J.H.}, \bibinfo{author}{Lee, J.H.}, \bibinfo{author}{Kang, H.},
  \bibinfo{author}{Park, L.J.}, \bibinfo{author}{Kim, D.J.},
  \bibinfo{author}{Yang, H.}, \bibinfo{author}{Hur, J.}, \bibinfo{author}{Kang,
  D.Y.}, et~al., \bibinfo{year}{2019}.
\newblock \bibinfo{title}{Deep learning segmentation of major vessels in x-ray
  coronary angiography}.
\newblock \bibinfo{journal}{Scientific reports} \bibinfo{volume}{9},
  \bibinfo{pages}{1--11}.
%Type = Misc
\bibitem[{Yoo(2020)}]{yoo2020deep}
\bibinfo{author}{Yoo, T.K.}, \bibinfo{year}{2020}.
\newblock \bibinfo{title}{Deep learning-based style transfer from
  ultra-widefield to traditional fundus photography}.
\newblock \URLprefix \url{https://data.mendeley.com/datasets/m3kg8p8cxf/2}.
%Type = Article
\bibitem[{Zhao et~al.(2019)Zhao, Chen, Hou and He}]{zhao2019segmentation}
\bibinfo{author}{Zhao, F.}, \bibinfo{author}{Chen, Y.}, \bibinfo{author}{Hou,
  Y.}, \bibinfo{author}{He, X.}, \bibinfo{year}{2019}.
\newblock \bibinfo{title}{Segmentation of blood vessels using rule-based and
  machine-learning-based methods: a review}.
\newblock \bibinfo{journal}{Multimedia Systems} \bibinfo{volume}{25},
  \bibinfo{pages}{109--118}.
%Type = Inproceedings
\bibitem[{Zhong et~al.(2021)Zhong, Yuan, Wu, Yuan, Peng and
  Wang}]{Zhong_2021_ICCV}
\bibinfo{author}{Zhong, Y.}, \bibinfo{author}{Yuan, B.}, \bibinfo{author}{Wu,
  H.}, \bibinfo{author}{Yuan, Z.}, \bibinfo{author}{Peng, J.},
  \bibinfo{author}{Wang, Y.X.}, \bibinfo{year}{2021}.
\newblock \bibinfo{title}{Pixel contrastive-consistent semi-supervised semantic
  segmentation}, in: \bibinfo{booktitle}{Proceedings of the IEEE/CVF
  International Conference on Computer Vision (ICCV)}, pp.
  \bibinfo{pages}{7273--7282}.

\end{thebibliography}

\clearpage
\section*{Supplementary Material}

\subsection*{Noise corruption study}

\begin{table}[h!]
    \caption{Quantitative results of the noise corruption study on the XCAD test dataset. For each Poisson and Gaussian noise, we use different levels of $\lambda$ and $\sigma$, respectively.}
    \centering
    \resizebox{\linewidth}{!}{
    \begin{tabular}{clccccc}

      \toprule
      \multirow{2}{*}{\bf{Level}} &  \multirow{2}{*}{\bf{Metric}} & \multicolumn{3}{c}{\textbf{w/ only CA data} } & \multicolumn{2}{c}{\bf{w/ multi-domain data }}    \\ 
         \cmidrule(l){3-5}\cmidrule(l){6-7}
        &  & DA & SSVS & DARL & DA & {C-DARL} \\ \midrule

        \multicolumn{7}{l}{Poisson noise ($\lambda$)} \\ \midrule
         \multirow{3}{*}{10} & IoU & $0.373_{\pm 0.072}$ & $0.344_{\pm 0.077}$ & $0.463_{\pm 0.077}$ & $0.269_{\pm 0.083}$ & $\bf0.490_{\pm 0.089}$  \\ 
         & Dice & $0.540_{\pm 0.080}$ & $0.507_{\pm 0.086}$ & $0.630_{\pm 0.073}$ & $0.417_{\pm 0.102}$ & $\bf0.652_{\pm 0.082}$  \\ 
         & Precision & $0.488_{\pm 0.106}$ & $0.562_{\pm 0.113}$ & $0.705_{\pm 0.114}$ & $0.411_{\pm 0.120}$ & $\bf0.752_{\pm 0.107}$  \\ \cmidrule(l){3-7}
         
        \multirow{3}{*}{20} & IoU & $0.331_{\pm 0.068}$ & $0.247_{\pm 0.068}$ & $0.448_{\pm 0.081}$ & $0.182_{\pm 0.067}$ & $\bf0.467_{\pm 0.095}$  \\ 
         & Dice & $0.494_{\pm 0.078}$ & $0.391_{\pm 0.086}$ & $0.614_{\pm 0.078}$ & $0.302_{\pm 0.095}$ & $\bf0.631_{\pm 0.090}$  \\
         & Precision & $0.425_{\pm 0.095}$ & $0.457_{\pm 0.116}$ & $0.709_{\pm 0.115}$ & $0.289_{\pm 0.102}$ & $\bf0.751_{\pm 0.109}$  \\ \cmidrule(l){3-7}
         
         \multirow{3}{*}{30} & IoU & $0.270_{\pm 0.070}$ & $0.151_{\pm 0.060}$ & $0.429_{\pm 0.086}$ & $0.119_{\pm 0.051}$ & $\bf0.441_{\pm 0.101}$  \\
         & Dice & $0.420_{\pm 0.086}$ & $0.257_{\pm 0.090}$ & $0.595_{\pm 0.085}$ & $0.210_{\pm 0.081}$ & $\bf0.605_{\pm 0.099}$  \\ 
         & Precision & $0.358_{\pm 0.092}$ & $0.313_{\pm 0.115}$ & $0.714_{\pm 0.117}$ & $0.201_{\pm 0.084}$ & $\bf0.752_{\pm 0.113}$ \\ 

        \midrule
        \multicolumn{7}{l}{Gaussian noise ($\sigma$)} \\ \midrule
         \multirow{3}{*}{10} & IoU & $0.353_{\pm 0.068}$ & $0.296_{\pm 0.076}$ & $0.452_{\pm 0.079}$ & $0.272_{\pm 0.080}$ & $\bf0.473_{\pm 0.093}$  \\ 
         & Dice & $0.518_{\pm 0.076}$ & $0.451_{\pm 0.091}$ & $0.618_{\pm 0.076}$ & $0.421_{\pm 0.097}$ & $\bf0.637_{\pm 0.088}$  \\ 
         & Precision & $0.465_{\pm 0.105}$ & $0.528_{\pm 0.118}$ & $0.710_{\pm 0.113}$ & $0.423_{\pm 0.116}$ & $\bf0.751_{\pm 0.109}$  \\ \cmidrule(l){3-7}
         
         \multirow{3}{*}{20} & IoU & $0.289_{\pm 0.069}$ & $0.164_{\pm 0.066}$ & $0.412_{\pm 0.087}$ & $0.193_{\pm 0.066}$ & $\bf0.426_{\pm 0.101}$  \\ 
         & Dice & $0.444_{\pm 0.085}$ & $0.276_{\pm 0.096}$ & $0.578_{\pm 0.087}$ & $0.318_{\pm 0.091}$ & $\bf0.590_{\pm 0.102}$  \\ 
         & Precision & $0.382_{\pm 0.098}$ & $0.360_{\pm 0.134}$ & $0.721_{\pm 0.117}$ & $0.318_{\pm 0.103}$ & $\bf0.752_{\pm 0.112}$  \\ \cmidrule(l){3-7}
         
         \multirow{3}{*}{30} & IoU & $0.221_{\pm 0.071}$ & $0.082_{\pm 0.046}$ & $0.365_{\pm 0.090}$ & $0.137_{\pm 0.056}$ & $\bf0.371_{\pm 0.111}$  \\ 
         & Dice & $0.356_{\pm 0.096}$ & $0.148_{\pm 0.077}$ & $0.528_{\pm 0.096}$ & $0.237_{\pm 0.086}$ & $\bf0.531_{\pm 0.123}$  \\ 
         & Precision & $0.312_{\pm 0.100}$ & $0.205_{\pm 0.109}$ & $0.728_{\pm 0.124}$ & $0.246_{\pm 0.097}$ & $\bf0.753_{\pm 0.130}$  \\ 
         
         %\cmidrule(l){2-8}
         % & \multirow{3}{*}{40} & IoU & $0.160_{\pm 0.069}$ & $0.045_{\pm 0.025}$ & $\bf0.318_{\pm 0.089}$ & $0.022_{\pm 0.019}$ & $0.304_{\pm 0.122}$  \\ 
         % & & Dice & $0.270_{\pm 0.103}$ & $0.085_{\pm 0.045}$ & $\bf0.475_{\pm 0.102}$ & $0.043_{\pm 0.035}$ & $0.452_{\pm 0.149}$  \\ 
         % & & Precision & $0.239_{\pm 0.101}$ & $0.121_{\pm 0.067}$ & $0.726_{\pm 0.137}$ & $0.068_{\pm 0.057}$ & $\bf0.748_{\pm 0.151}$  \\ \cmidrule(l){2-8}
         
         % & \multirow{3}{*}{50} & IoU & $0.102_{\pm 0.056}$ & $0.031_{\pm 0.017}$ & $\bf0.270_{\pm 0.083}$ & $0.017_{\pm 0.015}$ & $0.232_{\pm 0.126}$  \\ 
         % & & Dice & $0.180_{\pm 0.091}$ & $0.060_{\pm 0.031}$ & $\bf0.418_{\pm 0.103}$ & $0.033_{\pm 0.028}$ & $0.360_{\pm 0.168}$  \\ 
         % & & Precision & $0.169_{\pm 0.094}$ & $0.086_{\pm 0.046}$ & $\bf0.714_{\pm 0.154}$ & $0.048_{\pm 0.043}$ & $0.710_{\pm 0.197}$ \\ 
              
    \bottomrule     
    \end{tabular}
    }
    \label{tab:supple_noise}
    \vspace{-0.2cm}
\end{table}

\end{document}